 \documentclass[final,authoryear,5p,times,twocolumn]{elsarticle}
 
\usepackage{graphicx}
\usepackage{natbib}

\usepackage{amssymb,amsmath}

\journal{Planetary and Space Science}

\begin{document}

\begin{frontmatter}



\title{Properties of interstellar dust responsible for extinction laws 
with unusually \\ low total-to-selective extinction ratios of $R_V = 1$--2}


\author[label1]{Takaya Nozawa}
\ead{takaya.nozawa@nao.ac.jp}
\address[label1]{National Astronomical Observatory of Japan, Mitaka, 
Tokyo 181-8588, Japan}

\begin{abstract}

It is well known that the extinction properties along lines of sight to 
Type Ia supernovae (SNe Ia) are described by steep extinction curves 
with unusually low total-to-selective extinction ratios of 
$R_V =$ 1.0--2.0.
In order to reveal the properties of interstellar dust that causes 
such peculiar extinction laws, we perform the fitting calculations 
to the measured extinction curves by applying a two-component 
dust model composed of graphite and silicate.
As for the size distribution of grains, we consider two function 
forms of the power-law and lognormal distributions.
We find that the steep extinction curves derived from the 
one-parameter formula by \citet{car89} with $R_V = 2.0$, $1.5$, 
and $1.0$ can be reasonably explained even by the simple power-law 
dust model that has a fixed power index of $-3.5$ with the 
maximum cut-off radii of $a_{\rm max} \simeq$ 0.13 $\mu$m, 
0.094 $\mu$m, and 0.057 $\mu$m, respectively.
These maximum cut-off radii are smaller than 
$a_{\rm max} \simeq 0.24$ $\mu$m considered to be valid in the Milky 
Way, clearly demonstrating that the interstellar dust responsible for 
steep extinction curves is highly biased to smaller sizes.
We show that the lognomal size distribution can also lead to
good fits to the extinction curves with $R_V =$ 1.0--3.1 by 
taking the appropriate combinations of the relevant parameters.
We discuss that the extinction data at ultraviolet 
wavelengths are essential for constraining the composition and size 
distribution of interstellar dust.

\end{abstract}

\begin{keyword}
Dust \sep Extinction curves \sep Galaxy evolution \sep
Interstellar medium \sep Milky Way \sep Type Ia supernovae
\end{keyword}

\end{frontmatter}


\section{Introduction} 
\label{sec:intro}

The reddening of Type Ia supernovae (SNe Ia) by dust grains is one
of the largest uncertainties that limit the current precision of the 
cosmological parameters \citep[e.g.,][]{nor08}.
The reddening laws along lines of sight to SNe Ia are generally 
measured through the total-to-selective extinction ratio 
$R_V \equiv A_V /(A_B -A_V)$, where $A_V$ and $A_B$ are the 
extinction in $V$ and $B$ bands, respectively.
From the analyses of numerous samples of SNe Ia, many studies 
suggest that, to minimize the residual on the Hubble diagram, 
$R_V$ toward SNe Ia must be in a range of $R_V =$ 1.0--2.0 
\citep[e.g.,][]{con07, sul10, lam10}, which is considerably lower 
than the range $R_V =$ 2.2--5.5 measured in the Milky Way (MW).
The origin of such unusually low $R_V$ thus has been an important 
issue to be resolved for the applicability of SNe Ia as the cosmic 
standard candles.

The multi-band observations of individual SNe Ia also seem to 
commonly show the values as low as $R_V \le \sim 2.0$ 
\citep[][and reference therein]{how11}, although there is one study 
suggesting that weakly reddened SNe Ia have the values similar to the 
average $R_V = 3.1$ in the MW \citep{fol10}.
For these nearby SNe Ia, not only $R_V$ but also the extinction 
curves, namely the wavelength-dependence of extinction, can be 
obtained from optical to near-infrared photometries and spectra.
The derived extinction curves are much steeper than the MW average 
extinction curve and are nicely fitted with the one-parameter 
formula given in \citet*{car89} by taking $R_V \simeq$ 1.0--2.0.
Such non-standard extinction laws indicate that the properties of 
interstellar dust in host galaxies of SNe Ia are different from those in 
the MW or that other environmental effects associated with SNe Ia 
affect the apparent shapes of extinction curves.

One of the possible processes that have been suggested as provoking 
the steep extinction curves is the multiple scattering by dust grains 
surrounding SNe Ia;
it has been shown that multiple scattering of photons in a
circumstellar dust shell with a visual optical depth of $\tau \simeq 1.0$ 
can substantially steepen the extinction curve
\citep[][but see also Nagao, Maeda, \& Nozawa, 2016]{wan05, 
goo08, ama11}.
However, if there exist such a moderately optically thick dust shell, 
we also expect thermal emission from the circumstellar dust that is 
heated by the SN radiation (so-called infrared light echo).
\citet{joh13} observed three nearby SNe Ia with {\it Herschel} but 
did not detect any far-infrared emission.
\citet{joh14} also reported the non-detection with {\it Spitzer} at 
3.6 $\mu$m and 4.5 $\mu$m for several SNe Ia and placed the mass 
limit of $\le$ $\sim 10^{-5}$ $M_\odot$ on the amount of the 
circumstellar dust.
Furthermore, by comparing an infrared light echo model and near-infrared 
observations of SNe Ia samples, \citet{mae15} put the upper limits 
of $\tau_B \le \sim 0.1$ on the $B$-band optical depths of 
circumstellar dust shells.
These works point out that there would not be massive dust shells 
around SNe Ia so that the multiple scattering by local dust might not
be a valid explanation of the unusual extinction laws.

Recently, Type Ia SN 2014J was discovered in the starburst galaxy 
M82 at a distance of $\simeq$ 3.5 Mpc \citep{dal09}, which is the 
nearest among SNe Ia reported in the last thirty years.
This SN is highly reddened and thus offer the best opportunity to 
study the extinction property on its sightline.
The extensive observations have revealed that the extinction curve 
derived for SN 2014J is highly steep with a very low value of 
$R_V \simeq 1.5$ \citep[e.g.,][]{goo14, ama14, fol14, mar15}.
From a simultaneous fit to the extinction and polarization data, 
\citet{hoa15} claim that both interstellar and circumstellar dust 
are responsible for the anomalous extinction property toward SN 2014J.
However, the time invariability of the color \citep{bro15} and 
polarization \citep{kaw14, pat15} indicates that its peculiar extinction 
is mainly of interstellar-dust origin.
This seems to be also supported by non-detection of infrared excess, 
which otherwise would be caused by the circumstellar dust shell 
around SN 2014J \citep{joh14, tel15}.

These studies above provide an increasing number of evidence that 
the unusual extinction curves toward SNe Ia are likely to be originated 
by interstellar dust in their host galaxies.
Hence, the extinction curves measured for SNe Ia can be powerful 
tools for probing the properties of interstellar dust in external 
galaxies.
In general, a low $R_V$ is interpreted as corresponding to a smaller
average radius of dust than that in the MW.
\citet{gao15} searched for a physical model of relevant dust grains 
via the fitting to the measured color-excess curve of SN 2014J 
and found that their size distribution is biased to small radii 
relative to that in the MW.
However, it has not been systematically investigated how small the 
interstellar dust should be, to produce the steep extinction curves 
appeared for SNe Ia.
As stated above, the extinction laws toward SNe Ia are well 
described by the extinction curves derived from the empirical 
formula by \citet{car89} 
(which we, hereafter, refer to as the CCM curves).
Therefore, in this paper, we aim to comprehensively understand what 
composition and size distribution of interstellar dust can reproduce
the CCM curves with exceptionally low values of $R_V \le 2.0$.

The paper is organized as follows.
In Section \ref{sec:model}, the procedure of the fitting 
calculations and the model of dust used for deriving the extinction 
curves are described.
The results of the calculations are presented in Section 
\ref{sec:results} and discussed in Section \ref{sec:discussion}.
In Section \ref{sec:conclusion}, our main conclusions are remarked.
Throughout this paper, dust grains are assumed to be spherical.

\section{Fitting model of extinction curves} 
\label{sec:model}

\subsection{Data on extinction curves}
\label{subsec:extdata}

The main aim of this study is to explore how a variety of the CCM 
curves described by different $R_V$ can translate to the properties of 
interstellar dust.
In order to do this, we perform the calculations of fitting to the 
CCM curves by applying a simple dust model.
However, we do not try to fit the whole shape of the CCM curve by 
finely spacing wavelengths from 0.125 $\mu$m to 3.5 $\mu$m under 
consideration.
Indeed, the continuous CCM curves have been derived by interpolating 
the data of extinction obtained from ultraviolet (UV) spectra and 
optical to near-infrared photometries \citep{car89}.
Given that the extinction curves are usually extracted on the basis of
photometric observations, it is practical to consider the extinction at 
the specific wavelengths where the observational data are available.
In this study, we take, as the reference wavelengths, the effective 
wavelengths of widely-used optical to near-infrared photometric filters
and wide-band UV photometric filters onboard representative satellites 
such as {\it Habble Space Telescope (HST)}, {\it GALEX}, and 
{\it Swift}.
Table 1 presents the reference wavelengths $\lambda_i$ adopted in this 
study.
At these wavelengths, we calculate the extinction values 
$A_{\lambda_i}/A_V$ normalized to that in $V$ band for different $R_V$ 
by means of the CCM formula and use them as the data of extinction 
curves to be fitted.

\begin{table*}
\caption{Reference wavelengths adopted in this study.}
\label{tab:refwl}
\begin{center}
\begin{tabular}{@{}ccccc} \hline
Wavelength $\lambda_i$  &  $1/\lambda_i$  & $\sigma_i$\,$^a$ 
& Photometric filters & References\,$^b$ \\
($\mu$m)  & ($\mu$m$^{-1}$) &  &  & 
\\ \hline
0.125  &  8.000  & 0.32* & --- &  \\
0.1528 & 6.545 & 0.19* & FUV ({\it GALEX}) & (1) \\
0.1928 & 5.187 & 0.15* & uvw2 ({\it Swift}) & (2) \\
0.2224  & 4.496 & 0.2* & F218W ({\it HST}), NUV ({\it GALEX})\,$^c$ & (3), (1) \\
0.2359  & 4.239 & 0.2 & F225W ({\it HST}) & (3) \\
0.26   & 3.846 & 0.12* & uvw1 ({\it Swift}) & (2) \\
0.2704  & 3.698 & 0.12 & F275W ({\it HST}) & (3) \\
0.3355  & 2.981 & 0.065* & F336W ({\it HST}) & (3) \\
0.3531 & 2.832 & 0.065 & $u$ band & (4) \\
0.365  & 2.740  & 0.022* & $U$ band & \\
0.3921  & 2.550 & 0.022 & F390W ({\it HST}) & (3) \\
0.44   & 2.273 & 0.02* & $B$ band & \\
0.4627 & 2.161 & 0.02 & $g$ band & (4) \\
0.55   & 1.818 & --- & $V$ band & \\
0.614  & 1.629 & 0.02 & $r$ band & (4) \\
0.66   & 1.515 & 0.02* & $R_{\rm c}$ band & \\
0.7467 & 1.339 & 0.02 & $i$ band & (4) \\
0.81   & 1.235 & 0.027 & $I_{\rm c}$ band & \\
0.8887 & 1.125 & 0.027* & $z$ band & (4) \\
1.25  & 0.800 & 0.03* & $J$ band & (5) \\
1.65  & 0.606 & 0.034* & $H$ band & (5) \\
2.16  & 0.463 & 0.04* & $K_s$ band & (5) \\ 
3.4   & 0.294 & 0.06* & $L$ band & \\
\hline
\end{tabular}
\end{center}
$^a$The uncertainties of extinction $(A_{\lambda_i}/A_V)_{\rm CCM}$ 
obtained from the CCM formula at the reference wavelengths $\lambda_i$.
The values marked by asterisks are taken from \citet{car89}, while 
the others are deduced from those at the adjacent reference wavelengths.\\
$^b$References for the wavelengths: (1) \citet{mor05}, (2) \citet{poo08}, 
(3) \citet{dre16}, (4) 2.5 m reference in \citet{doi10}, 
(5) 2MASS bands in \citet{skr06}. \\
$^c$For the {\it GALEX} NUV band, the effective wavelength is 0.2271 $\mu$m.
\end{table*}

\subsection{Model of interstellar dust}\label{subsec:dustmodel}

For the model of interstellar dust, we adopt a two-component 
model consisting of graphite and silicate.
As for the size distribution of dust, we consider two simple function 
forms: one is the power-law size distribution given as
\begin{eqnarray}
n_j(a) = C_j f_j(a) = C_j a^{-q_j}, 
\label{eq:powerlaw}
\end{eqnarray}
where $n_j(a) da$ is the number density of grain species $j$ ($j$ 
denotes graphite or silicate) with radii between $a$ and $a + da$. 
The normalization factor $C_j$ is related to the specific mass of dust 
grains, $m_j$, as
\begin{eqnarray}
C_j = m_j \left( \frac{4 \pi \rho_j}{3} \int_{a_{{\rm min}, j}}^{a_{{\rm max}, j}} 
a^3 f_j(a) da \right)^{-1} = \frac{m_j}{X_j},
\end{eqnarray}
with $\rho_j$ being the material density ($\rho_{\rm gra} = 2.24$ 
g cm$^{-3}$ and $\rho_{\rm sil} = 3.3$ g cm$^{-3}$), 
$a_{{\rm max}, j}$ the maximum cut-off radius, and $a_{{\rm min}, j}$ 
the minimum cut-off radius of grain species $j$.
It is well known that, for this grain composition and size distribution 
function, the average extinction curve in the MW is nicely reproduced 
by taking $q_{\rm gra} = q_{\rm sil} = 3.5$,  
$a_{\rm max, gra} = a_{\rm max, sil} \simeq 0.25$ $\mu$m,
$a_{\rm min, gra} = a_{\rm min, sil} \simeq 0.005$ $\mu$m,
which has been referred to as the MRN dust model \citep*{mat77, dra84}.
Note that the power-law distribution is originated from the collisional
fragmentation of dust grains \citep{bie80}, which is considered as one of 
the main physical processes that modulate the size distribution of 
interstellar dust \citep[e.g.,][]{hir13, asa13}.

On the other hand, the size distributions of dust produced in stellar 
sources would not necessarily follow the power-law distribution;
it has been suggested that the size distributions of dust ejected from 
core-collapse supernovae (CCSNe) and asymptotic giant branch (AGB) 
stars are lognormal-like with peaks around 0.1--1.0 $\mu$m 
\citep[e.g.,][]{noz07, yas12}.
Hence, as the other size distribution function, we adopt the lognormal 
form of
\begin{eqnarray}
n_j(a) = C_j f_j(a) = 
\frac{C_j}{\sqrt{2 \pi} a \gamma_j} 
\exp \left[ - \frac{(\ln a -\ln a_{0, j} )^2}{2 \gamma_j^2} \right], 
\label{eq:lognormal}
\end{eqnarray}
where $a_{0, j}$ and $\gamma_j$ are the characteristic grain radius and  
the standard deviation of the lognormal distribution, respectively.
The normalization factor $C_j$ in Equation (3) is also determined 
from Equation (2), for which we fix $a_{{\rm max}, j} = 10$ $\mu$m and 
$a_{{\rm min}, j} = 5 \times 10^{-4}$ $\mu$m.

There are some elaborate interstellar dust models that consider various 
grain components and more complicated functional forms for grain size 
distributions \citep*[e.g.,][]{wei01, cla03, zub04, jon13}.
These dust models are tailored to produce excellent fits to the average 
MW extinction curve and to meet the constraints on elemental abundances 
in the interstellar medium.
However, the information on elemental abundances is poorly
available in general, especially for host galaxies of SNe Ia.
Thus, we do not take into account the abundance constraints in the 
present analysis and address only the reproduction of 
wavelength dependence of extinction.
Furthermore, our goal is to grasp the systematic behavior of how the 
properties of dust vary according to the extinction curves, not to seek 
a unique set of the grain composition and size distribution that gives the 
best fit to each extinction curve.
For this purpose, the simple dust models as proposed here are favorable, 
and they may also be useful to illustrate the applicability of the power-law 
and lognormal size distributions.

Under the assumption that both graphite and silicate grains are uniformly 
distributed in interstellar space, the extinction curve $A_{\lambda}/A_V$ 
is calculated, for the dust model described above, as
\begin{eqnarray}
\frac{A_{\lambda}}{A_V} = \frac{\sum_j K_j \int_{a_{{\rm min}, j}}^{a_{{\rm max}, j}} 
\pi a^2 f_j(a) Q_{\lambda, j}^{\rm ext}(a) da}{\sum_j K_j 
\int_{a_{{\rm min}, j}}^{a_{{\rm max}, j}} \pi a^2 f_j(a) Q_{V, j}^{\rm ext}(a) da},
\label{eq:modelext}
\end{eqnarray}
where
\begin{eqnarray}
K_{\rm gra} = \frac{C_{\rm gra}}{C_{\rm sil}} 
= f_{\rm gs} \left( \frac{X_{\rm sil}}{X_{\rm gra}} \right) 
~~~ {\rm and} ~~~ K_{\rm sil} = 1
\end{eqnarray}
with the mass ratio $f_{\rm gs} = m_{\rm gra}/m_{\rm sil}$ of graphite to 
silicate.
The extinction coefficients $Q_{\lambda, j}^{\rm ext}(a)$ are calculated 
with Mie scattering from dielectric constants for graphite and astronomical
silicate (which we refer to simply as silicate in what follows) in 
\citet{dra84} and \citet{dra03}.
In computing $Q_{\lambda, j}^{\rm ext}$ for graphite, the $1/3$--$2/3$ 
approximation is employed to take the anisotropy into consideration
\citep{dra93}.

For the power-law size distribution, there are seven adjustable 
parameters:
$f_{\rm gs}$, which prescribes the mixture of graphite and silicate grains, 
and $q_{\rm gra}$, $ q_{\rm sil}$, $a_{\rm max, gra}$, 
$a_{\rm max, sil}$, $a_{\rm min, gra}$, and $a_{\rm min, sil}$,
which regulate the size distribution of each grain species.
As discussed in \citet{noz13},  $a_{{\rm min}, j}$ cannot be constrained 
very much unless the extinction data at wavelengths shorter than 
$\simeq$0.1 $\mu$m are used.
Therefore, we fix $a_{{\rm min}, j} = 0.005$ $\mu$m for both graphite 
and silicate.
For this power-law distribution, we consider five dust models, 
depending on the combinations of the free parameters and fixed 
parameters, as summarized in Table \ref{tab:dustmodelpl}.
For example, the dust model in which graphite and silicate have the 
same size distribution with 
$q = q_{\rm gra} = q_{\rm sil} = 3.5$ and 
$a_{\rm max} = a_{\rm max, gra} = a_{\rm max, sil}$ is referred to as 
Model 1.
On the other hand, the dust model in which all of $f_{\rm gs}$, 
$q_{\rm gra}$, $ q_{\rm sil}$, $a_{\rm max, gra}$,  and $a_{\rm max, sil}$ 
are treated as free parameters is referred to as Model 5.
Hereafter, we mainly show the results of the fitting calculations 
for Model 1.
This simple dust model reduces the number of parameters to only 
two ($f_{\rm gs}$ and $a_{\rm max}$) and can be regarded as a 
natural extention of the MRN dust model.

For the lognormal size distribution, the adjustable parameters are 
$f_{\rm gs}$, $a_{\rm 0, gra}$, $ a_{\rm 0, sil}$, 
$\gamma_{\rm gra}$, and $\gamma_{\rm sil}$.
For this size distribution, two dust models are considered, according 
to the numbers of the free parameters.
The details of the dust models for the lognormal size distribution 
are given in Table \ref{tab:dustmodelln} and are described in 
Section \ref{sec:lognormal}.

\subsection{Fitting procedure} 
\label{subsec:fitting}

We carry out the fitting through comparison between the extinction 
data and the calculated extinction 
$y_{{\rm cal}, i} = (A_{\lambda_i}/A_V)_{\rm cal}$ 
at each reference wavelength $\lambda_i$.
The goodness of the fitting is evaluated by $\chi^2$, which might be,
for instance, given as
\begin{eqnarray}
\chi_0^2 = \frac{\langle \sigma \rangle^2}{N_{\rm data} - N_{\rm para}} 
\sum_i^{N_{\rm data}} 
\frac{ \left( y_{{\rm CCM}, i} - y_{{\rm cal}, i} \right)^2}{\sigma_i^2},
\label{eq:dispers1}
\end{eqnarray}
where $y_{{\rm CCM}, i} = (A_{\lambda_i}/A_V)_{\rm CCM}$ is the 
extinction derived from the CCM formula at $\lambda_i$, 
$N_{\rm data}$ is the number of the data to be fitted
(i.e., the number of the reference wavelengths under consideration
except for $V$ band),
$N_{\rm para}$ is the number of free parameters, 
$\sigma_i$ are weights, and 
$\langle \sigma \rangle^2 
= [ (1 / N_{\rm data}) \sum_i^{N_{\rm data}} (1 / \sigma_i^2) ]^{-1}$
is the adjustment coefficient for alleviating the large external errors 
arising from $\sigma_i^2$.
As for $\sigma_i$, which are generally assigned as the uncertainties of 
observational data \citep[e.g.,][]{gao15}, we take the standard 
deviation of $A_{\lambda}/A_V$ given in Table 2 of \citet{car89}.
For $\sigma_i$ not given in \citet{car89}, we infer their values from
those at the closest reference wavelengths.
The adopted values of $\sigma_i$ are provided in Table \ref{tab:refwl}.

It should be noted that a set of the best-fit parameters that minimizes 
$\chi_0^2$ in Equation (\ref{eq:dispers1}) does not necessarily 
produce apparently good fits to the entire range of extinction curves;
since $\sigma_i$ are considerably large at UV wavelengths 
(see Table \ref{tab:refwl}), the data of UV extinction are unimportantly 
treated in the fitting calculations, resulting in poor fits at UV 
wavelengths.
Therefore, as in some previous studies \citep[e.g.,][]{wei01}, we adopt 
an identical weight $\sigma_i = 1$, independent of $\lambda_i$, and 
assess the dispersion by
\begin{eqnarray}
\chi_1^2 = \frac{1}{N_{\rm data} - N_{\rm para}} 
\sum_i^{N_{\rm data}} \left( y_{{\rm CCM}, i} - y_{{\rm cal}, i} \right)^2.
\label{eq:dispers2}
\end{eqnarray}
In the case that the extinction data at all the reference wavelengths 
covering UV to near-infrared are exploited, $N_{\rm data} = 22$.
In Section \ref{sec:nonuv}, we will also show the results of the fitting 
calculations in the cases that the extinction data at UV wavelengths
are not taken into account.

\begin{table*}
\caption{Dust models with power-law size distributions and their 
parameter sets.}
\label{tab:dustmodelpl}
\begin{center}
\begin{tabular}{@{}cccc} \hline
Dust model & $N_{\rm para}$  & free parameters & constraints$\,^a$
\\ \hline 
Model 1 & 2 & $a_{\rm max}$, $f_{\rm gs}$ & $q_{\rm gra} = q_{\rm sil} = 3.5$,
$a_{\rm max} = a_{\rm max, gra} = a_{\rm max, sil}$  \\
Model 2 & 2 & $q$, $f_{\rm gs}$ & 
$q = q_{\rm gra} = q_{\rm sil}$, 
$a_{\rm max, gra} = a_{\rm max, sil} = 0.25$ $\mu$m \\
Model 3 & 3 & $a_{\rm max, gra}$, $a_{\rm max, sil}$, $f_{\rm gs}$ & 
$q_{\rm gra} = q_{\rm sil} = 3.5$ \\
Model 4 & 3 & $q$, $a_{\rm max,}$, $f_{\rm gs}$ & 
$q = q_{\rm gra} = q_{\rm sil}$,
$a_{\rm max} = a_{\rm max, gra} = a_{\rm max, sil}$ \\
Model 5 & 5 & $q_{\rm gra}$, $q_{\rm sil}$, $a_{\rm max, gra}$, $a_{\rm max, sil}$, 
$f_{\rm gs}$ & --- \\
\hline
\end{tabular}
\end{center}
$^a$For all of the models here, the minimum cut-off radii are fixed as
$a_{\rm min, gra} = a_{\rm min, sil} = 0.005$ $\mu$m.
\end{table*}

\begin{table*}
\caption{Dust models with lognormal size distributions and their 
parameter sets.}
\label{tab:dustmodelln}
\begin{center}
\begin{tabular}{@{}cccc} \hline
Dust model & $N_{\rm para}$  & free parameters & constraints$\,^a$
\\ \hline 
Model 1s & 3 & $a_{\rm 0, gra}$, $a_{\rm 0, sil}$, $f_{\rm gs}$ & 
$\gamma_{\rm gra} = \gamma_{\rm sil} = 0.5$ \\
Model 5s & 5 & $a_{\rm 0, gra}$, $a_{\rm 0, sil}$, $\gamma_{\rm gra}$,
$\gamma_{\rm sil}$, $f_{\rm gs}$ & --- \\
\hline
\end{tabular}
\end{center}
$^a$For the models with lognormal size distributions, the maximum and 
minimum cut-off radii are fixed to be 
$a_{\rm max, gra} = a_{\rm max, sil} = 10$ $\mu$m and
$a_{\rm min, gra} = a_{\rm min, sil} = 5 \times 10^{-4}$ $\mu$m, respectively.
\end{table*}

\begin{table*}
\caption{A set of the best-fit parameters obtained for dust models with
power-law size distributions.}
\label{tab:fitres}
\begin{center}
\begin{tabular}{@{}lccccccc} \hline
Dust model$\,^a$ & $q_{\rm gra}$ & $a_{\rm max, gra}$ & $q_{\rm sil}$ & 
$a_{\rm max, sil}$ & $f_{\rm gs}$ & $\chi_1$ & $R_V^{\rm cal}$ \\
 & & ($\mu$m) &  & ($\mu$m) & & & 
\\ \hline \hline
\multicolumn{8}{c}{$R_V^{\rm CCM} = 3.1$}  \\ \hline
Model 1 & 3.50 & 0.243 & 3.50 & 0.243 & 0.57 & 0.0464 & 3.38 \\
Model 2 & 3.53 & 0.250 & 3.53 & 0.250 & 0.60 & 0.0430 & 3.46 \\
Model 3 & 3.50 & 0.119 & 3.50 & 0.544 & 0.25 & 0.0462 & 3.44 \\
Model 4 & 3.58 & 0.274 & 3.58 & 0.274 & 0.60 & 0.0413 & 3.56 \\
Model 5 & 3.78 & 0.471 & 3.42 & 0.290 & 0.44 & 0.0358 & 3.46 \\ 
\hline \hline
\multicolumn{8}{c}{$R_V^{\rm CCM} = 2.0$}  \\ \hline
Model 1 & 3.50 & 0.134 & 3.50 & 0.134 & 0.46 & 0.0932 & 2.16 \\
Model 2 & 4.05 & 0.250 & 4.05 & 0.250 & 0.50 & 0.0893 & 2.66 \\
Model 3 & 3.50 & 0.0386 & 3.50 & 0.270 & 0.16 & 0.0550 & 2.41 \\
Model 4 & 3.84 & 0.174 & 3.84 & 0.174 & 0.49 & 0.0540 & 2.41 \\
Model 5 & 4.04 & 0.164 & 3.76 & 0.230 & 0.35 & 0.0368 & 2.34 \\ 
\hline \hline
\multicolumn{8}{c}{$R_V^{\rm CCM} = 1.5$}  \\ \hline
Model 1 & 3.50 & 0.0944 & 3.50 & 0.0944 & 0.49 & 0.0707 & 1.75 \\
Model 2 & 4.40 & 0.250 & 4.40 & 0.250 & 0.47 & 0.231 & 2.34 \\
Model 3 & 3.50 & 0.0884 & 3.50 & 0.0735 & 0.64 & 0.0535 & 1.72 \\
Model 4 & 3.67 & 0.102 & 3.67 & 0.102 & 0.48 & 0.0643 & 1.89 \\
Model 5 & 4.10 & 0.0903 & 3.85 & 0.200 & 0.31 & 0.0465 & 1.74 \\ 
\hline \hline
\multicolumn{8}{c}{$R_V^{\rm CCM} = 1.0$}  \\ \hline
Model 1 & 3.50 & 0.0572 & 3.50 & 0.0572 & 0.60 & 0.223 & 1.49 \\
Model 2 & 5.08 & 0.250 & 5.08 & 0.250 & 0.42 & 0.663 & 2.10 \\
Model 3 & 3.50 & 0.0627 & 3.50 & 0.0844 & 0.42 & 0.171 & 1.39 \\
Model 4 & 3.28 & 0.0546 & 3.28 & 0.0546 & 0.63 & 0.218 & 1.47 \\
Model 5 & 3.86 & 0.0600 & 3.77 & 0.128 & 0.30 & 0.158 & 1.34 \\ 
\hline
\end{tabular}
\end{center}
$^a$The free parameters and constraints adopted in the power-law 
dust models are described in Table \ref{tab:dustmodelpl}. 
\end{table*}

\section{Results of fitting calculations}
\label{sec:results}

\subsection{Extinction curves from dust models with power-law size distributions}
\label{sec:powerlaw}

We first consider the average MW extinction curve with 
$R_V^{\rm CCM} = 3.1$ to check if our fitting method yields the 
plausible results.
As mentioned above, this extinction curve can be well matched by the 
MRN dust model that is represented by the power-law size distribution 
with $q = 3.5$ and $a_{\rm max} = 0.25$ $\mu$m 
\citep{mat77, dra84, noz13}.
Thus, our fitting calculations should lead to the similar values for $q$ 
and $a_{\rm max}$.

The results of the fitting calculations are given in Table \ref{tab:fitres},
which summarizes a set of parameters that give the best fit for each 
dust model with a power-law size distribution.
For Model 1, which assumes $q = 3.5$ and 
$a_{\rm max} = a_{\rm max, gra} = a_{\rm max, sil}$, the best fit is 
achieved with $a_{\rm max} = 0.243$ $\mu$m and $f_{\rm gs} = 0.57$.
On the other hand, if we fix as $a_{\rm max} = 0.25$ $\mu$m 
(Model 2), the power index $q = 3.53$ offers the best fit with
$f_{\rm gs} = 0.60$.
Hence, our simplest dust models (Models 1 and 2) present the results 
consistent with the MRN dust model, demonstrating that our fitting 
method works well.

\begin{figure*}
\begin{tabular}{cc}
\begin{minipage}[t]{0.45\hsize}
\includegraphics[width=1\textwidth]{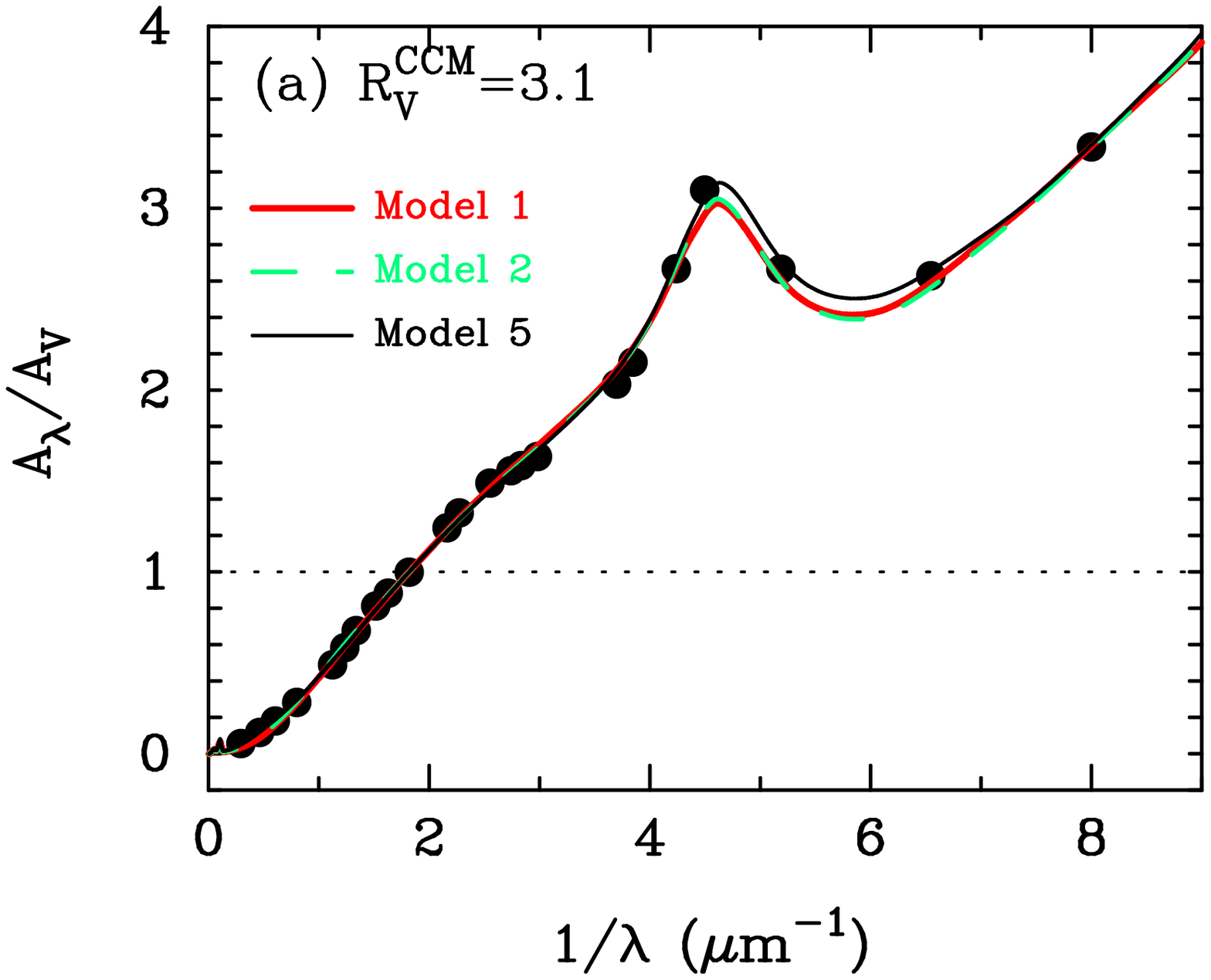}
\end{minipage} &
\begin{minipage}[t]{0.45\hsize}
\includegraphics[width=1\textwidth]{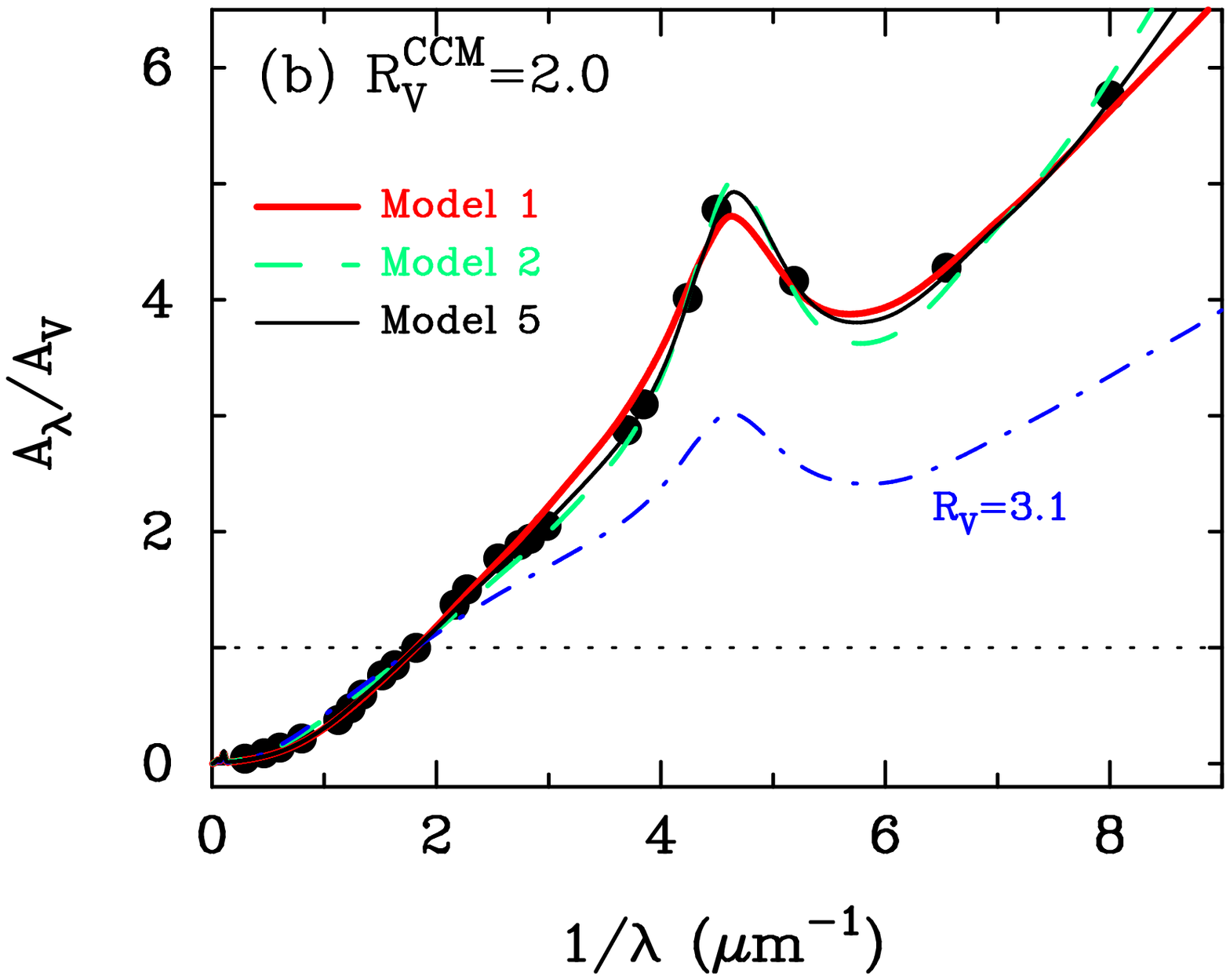}
\end{minipage}
\end{tabular}
\begin{tabular}{cc}
\begin{minipage}[t]{0.45\hsize}
\includegraphics[width=1\textwidth]{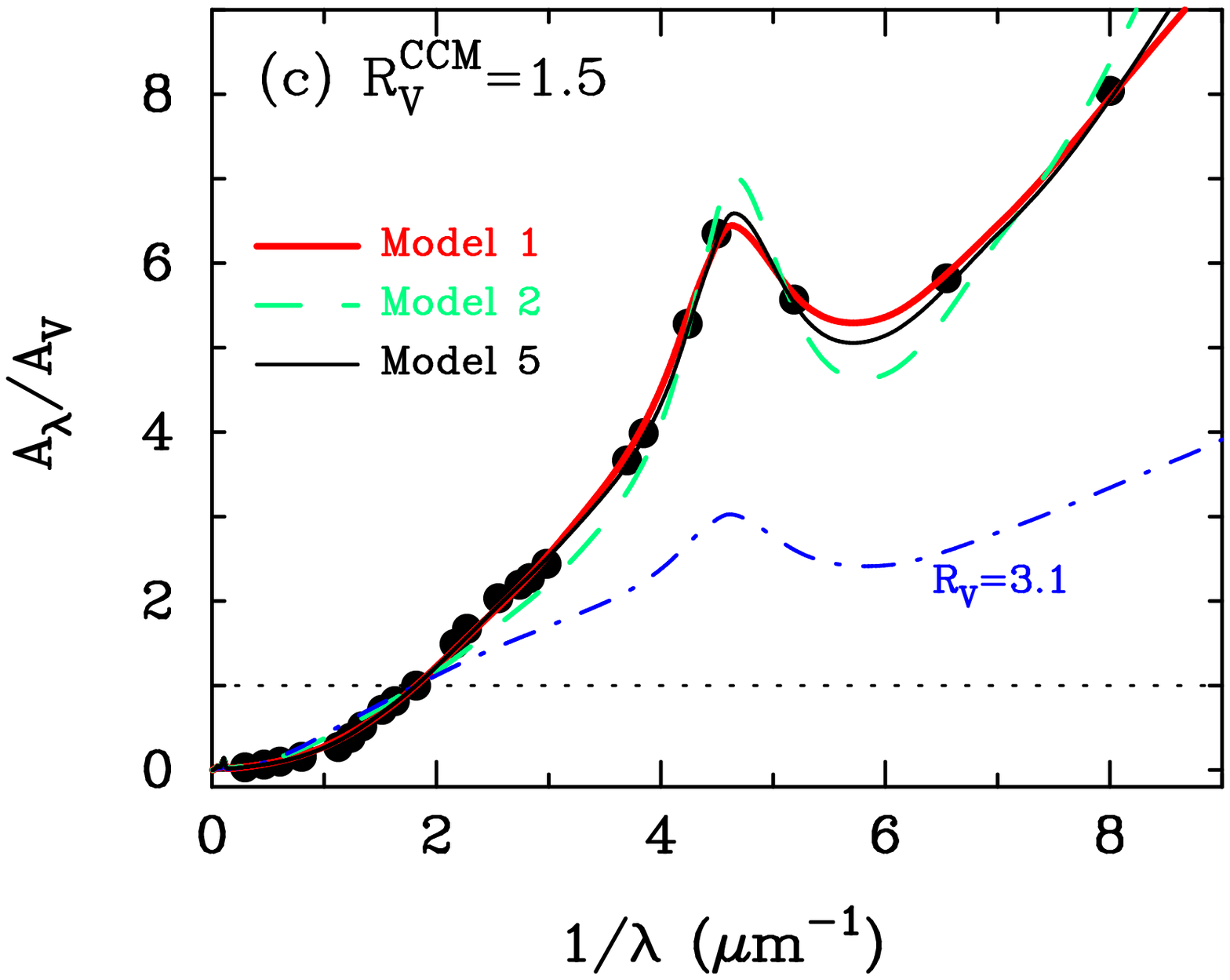}
\end{minipage} &
\begin{minipage}[t]{0.45\hsize}
\includegraphics[width=1\textwidth]{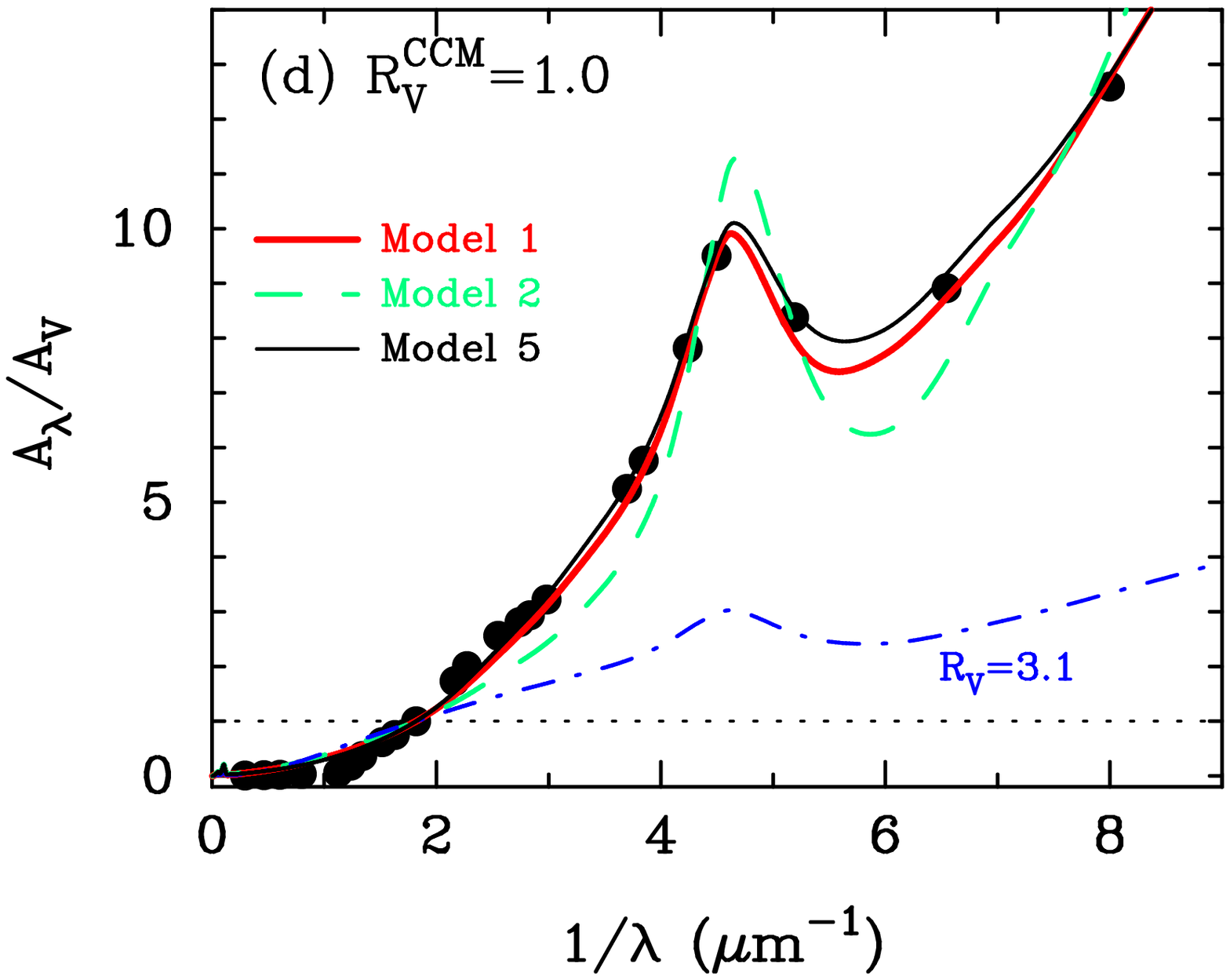}
\end{minipage}
\end{tabular}
\caption{Extinction curves calculated from the best-fit parameters 
in Model 1 (thick solid), Model 2 (dashed), and Model 5 (thin solid)
with power-law size distributions for (a) $R_V^{\rm CCM} = 3.1$ 
(b) $2.0$, (c) $1.5$, and (d) $1.0$.
The filled circles are the extinction data at the reference wavelengths,
derived from the CCM formula for each $R_V^{\rm CCM}$ value.
For the panels (b)--(d), the extinction curve derived from Model 1 for 
$R_V^{\rm CCM} = 3.1$ is shown by the dot-dashed line for reference.
\label{fig:powerlaw}}
\end{figure*}

The extinction curves derived from these best-fit parameters are 
shown in Figure \ref{fig:powerlaw}(a).
We can see that the extinction curves obtained from Model 1 and Model 2
are fully overlapped and reproduce the whole shape of the CCM curve 
with $R_V^{\rm CCM} = 3.1$.
In Figure \ref{fig:powerlaw}(a), we also depict the extinction curve from 
Model 5.
Since Model 5 treats the five quantities ($q_{\rm gra}$, 
$a_{\rm max, gra}$, $q_{\rm sil}$, $a_{\rm max, sil}$, and $f_{\rm gs}$) 
as free parameters, the best-fit values result in the least dispersion 
among the dust models considered for the power-law size distribution
(see Table \ref{tab:fitres}).
However, as is observed from Figure \ref{fig:powerlaw}(a), the 
match between the resulting extinction curve and the extinction data 
is not largely superior to that for Model 1.
Given that Model 1 (and Model 2) gives a satisfactory fit despite its 
small number of free parameters, such a simplest dust model should be 
regarded as being instructive for understanding the properties of dust 
in comparison with the MRN model.

Next we search for the dust models that can account for very steep 
extinction curves with $R_V^{\rm CCM} \le 2.0$.
Figures \ref{fig:powerlaw}(b), \ref{fig:powerlaw}(c), and 
\ref{fig:powerlaw}(d) show the results of the fitting to the CCM 
curves with $R_V^{\rm CCM} = 2.0$, $1.5$, and $1.0$, respectively.
For Model 1 with $q = 3.5$ and $a_{\rm max, gra} = a_{\rm max, sil}$, 
the least dispersions are achieved with
$a_{\rm max} = 0.134$ $\mu$m, $0.0944$ $\mu$m,
and $0.0572$ $\mu$m for $R_V^{\rm CCM} = 2.0$, $1.5$, and $1.0$.
For these cases, the mass ratio of graphite to silicate is in a range 
of $f_{\rm gs} =$ 0.45--0.6.
The dispersions are considerably small for $R_V^{\rm CCM} = 2.0$ 
and $1.5$, demonstrating that the fits are adequate.
The fit to the $R_V^{\rm CCM} = 1.0$ curve is not good very much 
at optical to near-infrared wavelengths, but the calculated extinction 
curve can roughly reproduce the entire wavelength-dependence of 
extinction.

In Figures \ref{fig:powerlaw}(b)--\ref{fig:powerlaw}(d), we also plot 
the extinction curves calculated from the best-fit combination of 
$f_{\rm gas}$ and $q$ for Model 2.
When the maximum cut-off radii are fixed as $a_{\rm max} = 0.25$ 
$\mu$m, the optimum values of $q$ are 4.05, 4.40, and 5.08, 
respectively, for $R_V^{\rm CCM} = 2.0$, $1.5$, and $1.0$.
This implies that as the extinction curves become steeper, the 
steeper size distributions are required.
However, as is obvious from Figures \ref{fig:powerlaw}(c) and 
\ref{fig:powerlaw}(d), the best-fit parameters for Model 2 lead to 
a poor match to the CCM curves with $R_V^{\rm CCM} = 1.5$ and 
$1.0$.
Therefore, only enhancement in $q$, with a fixed $a_{\rm max}$, 
is not sufficient for describing the highly steep extinction curves 
with $R_V^{\rm CCM} \le 1.5$.
Nevertheless, for $R_V^{\rm CCM} = 2.0$, the size distribution 
from Model 2 with $q = 4.05$ and $a_{\rm max} = 0.25$ $\mu$m 
yields a better fit than that for Model 1 with $q = 3.5$ and 
$a_{\rm max} = 0.134$ $\mu$m.

These results indicate that the steeper extinction curves represented 
by the values as low as $R_V^{\rm CCM} =$ 1.0--2.0 can be explained 
by the dust model whose power-law size distribution is skewed to 
smaller radii than the MRN model.
More specifically, if $a_{\rm max} = 0.25$ $\mu$m is held, the 
power index is needed to be increased from $q \simeq 3.5$ for
$R_V^{\rm CCM} =$ 3.1 up to $q \simeq$ 4.0 for 
$R_V^{\rm CCM} =$ 2.0.
On the other hand, for $R_V^{\rm CCM} =$ 1.0--2.0, 
a better fit can be generally obtained through reducing the maximum 
cut-off radius by a factor of 2--5 in comparion to the MRN model, 
with a fixed index of $q = 3.5$ and a mass ratio of 
$f_{\rm gs} =$ 0.45--0.6.
Given the peculiality of the extinction curves as never appeared 
in the MW, the maximum cut-off radii of $a_{\rm max} \simeq$ 
0.06--0.13 $\mu$m are not too small and thus may not be unrealistic.
The important consequence of the present analysis is that the 
remarkably tilted extinction curves measured for SNe Ia can be 
described in the context of the simple dust model with the power-law 
size distribution.

\begin{table*}
\caption{A set of the best-fit parameters obtained for dust models with
lognormal size distributions.}
\label{tab:fitres1}
\begin{center}
\begin{tabular}{@{}lccccccc} \hline
Dust model$\,^a$ & $\gamma_{\rm gra}$ & $a_{\rm 0, gra}$ & $\gamma_{\rm sil}$ & 
$a_{\rm 0, sil}$ & $f_{\rm gs}$ & $\chi_1$ & $R_V^{\rm cal}$ \\
 & & ($\mu$m) &  & ($\mu$m) & & & 
\\ \hline \hline
\multicolumn{8}{c}{$R_V^{\rm CCM} = 3.1$}  \\ \hline
Model 1s & 0.50 & 0.0358 & 0.50 & 0.0253 & 0.91 & 0.139 & 2.81 \\
Model 5s & 1.32 & $5.05 \times 10^{-4}$ & 1.34 & $5.12 \times 10^{-4}$ 
& 0.55 & 0.0675 & 3.56 \\ 
\hline \hline
\multicolumn{8}{c}{$R_V^{\rm CCM} = 2.0$}  \\ \hline
Model 1s & 0.50 & 0.0238 & 0.50 & 0.0201 & 0.78 & 0.208 & 1.87 \\
Model 5s & 1.18 & $5.60 \times 10^{-4}$ & 1.25 & $5.00 \times 10^{-4}$ 
& 0.42 & 0.0608 & 2.48 \\ 
\hline \hline
\multicolumn{8}{c}{$R_V^{\rm CCM} = 1.5$}  \\ \hline
Model 1s & 0.50 & 0.0186 & 0.50 & 0.0185 & 0.69 & 0.220 & 1.61 \\
Model 5s & 0.86 & $2.84 \times 10^{-3}$ & 1.06 & $1.48 \times 10^{-3}$ 
& 0.38 & 0.0890 & 1.95 \\ 
\hline \hline
\multicolumn{8}{c}{$R_V^{\rm CCM} = 1.0$}  \\ \hline
Model 1s & 0.50 & 0.0127 & 0.50 & 0.0172 & 0.54 & 0.239 & 1.49 \\
Model 5s & 0.50 & 0.0114 & 0.71 & $8.63 \times 10^{-3}$ & 0.39 & 0.232 & 1.50 \\ 
\hline
\end{tabular}
\end{center}
$^a$The free parameters and constraints adopted in the lognormal dust 
models are described in Table \ref{tab:dustmodelln}. 
\end{table*}

\begin{figure*}
\begin{tabular}{cc}
\begin{minipage}[t]{0.45\hsize}
\includegraphics[width=1\textwidth]{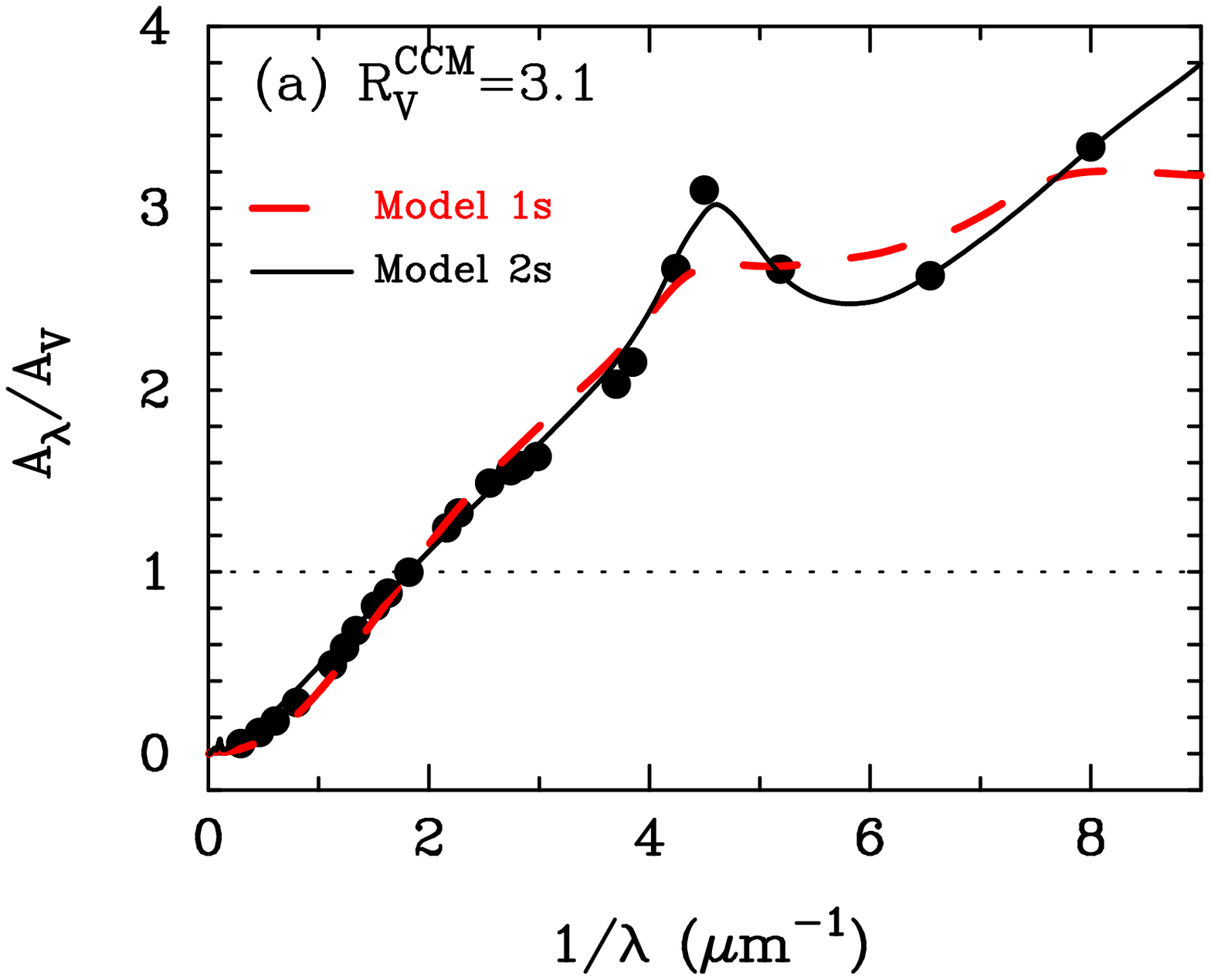}
\end{minipage} &
\begin{minipage}[t]{0.45\hsize}
\includegraphics[width=1\textwidth]{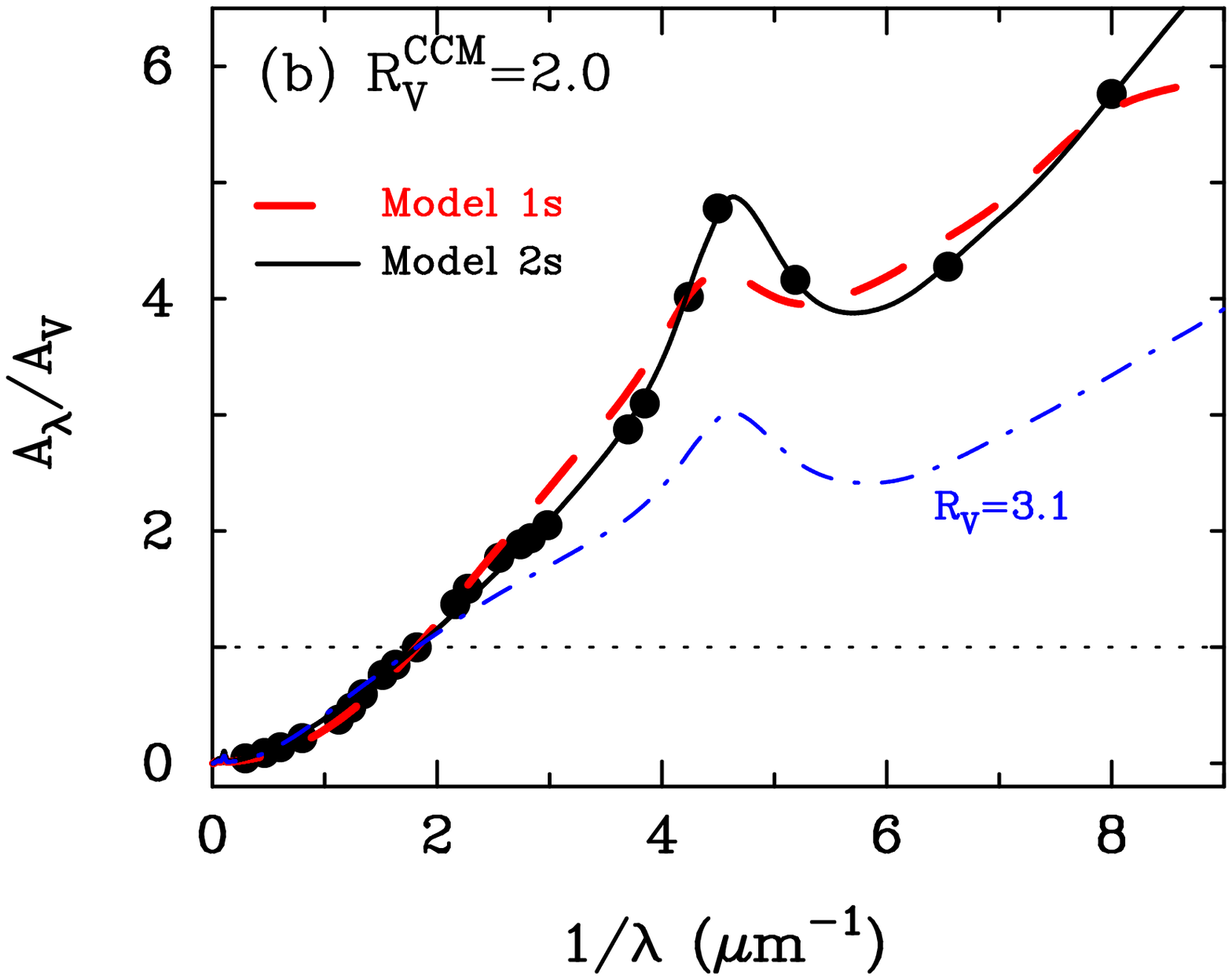}
\end{minipage}
\end{tabular}
\begin{tabular}{cc}
\begin{minipage}[t]{0.45\hsize}
\includegraphics[width=1\textwidth]{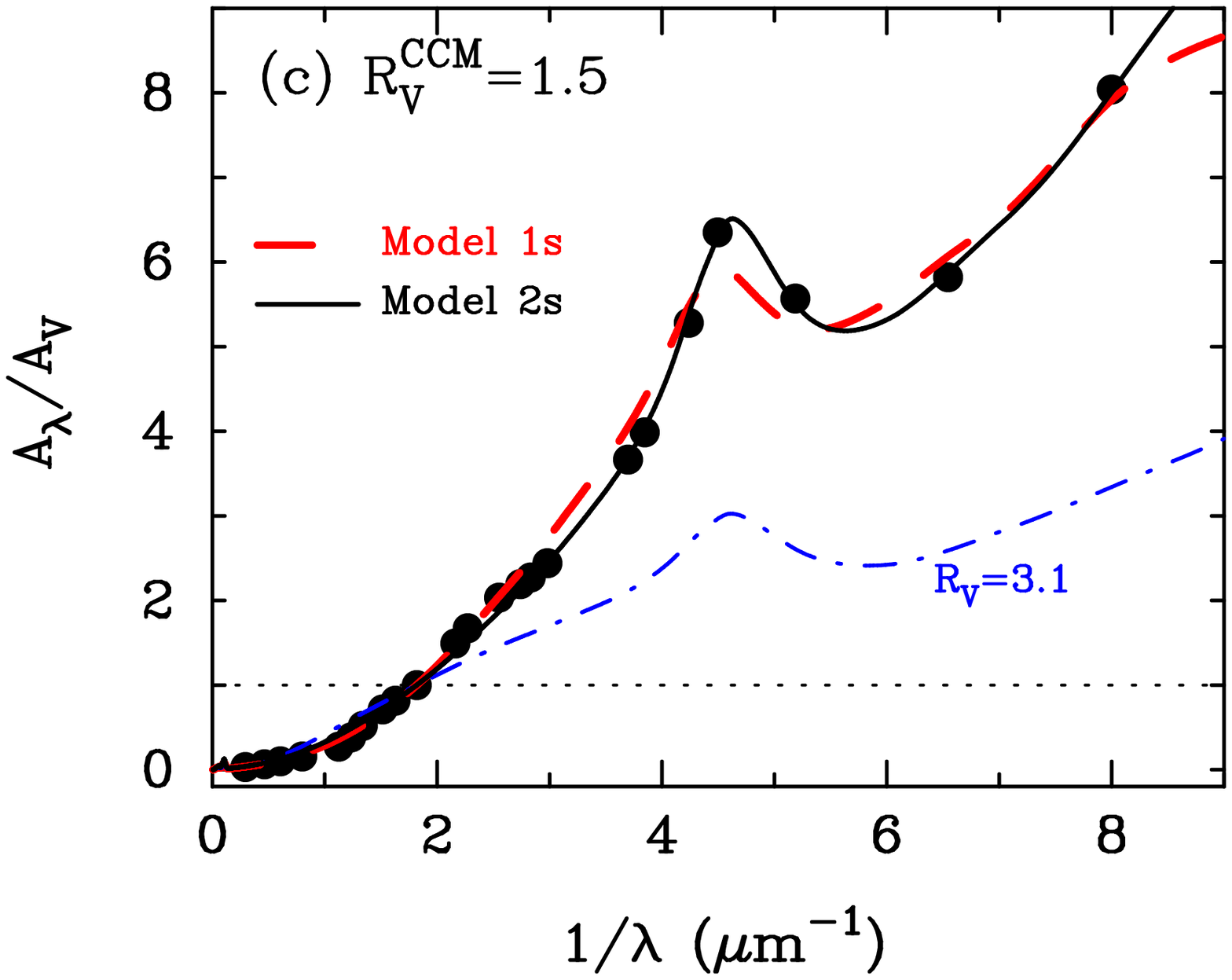}
\end{minipage} &
\begin{minipage}[t]{0.45\hsize}
\includegraphics[width=1\textwidth]{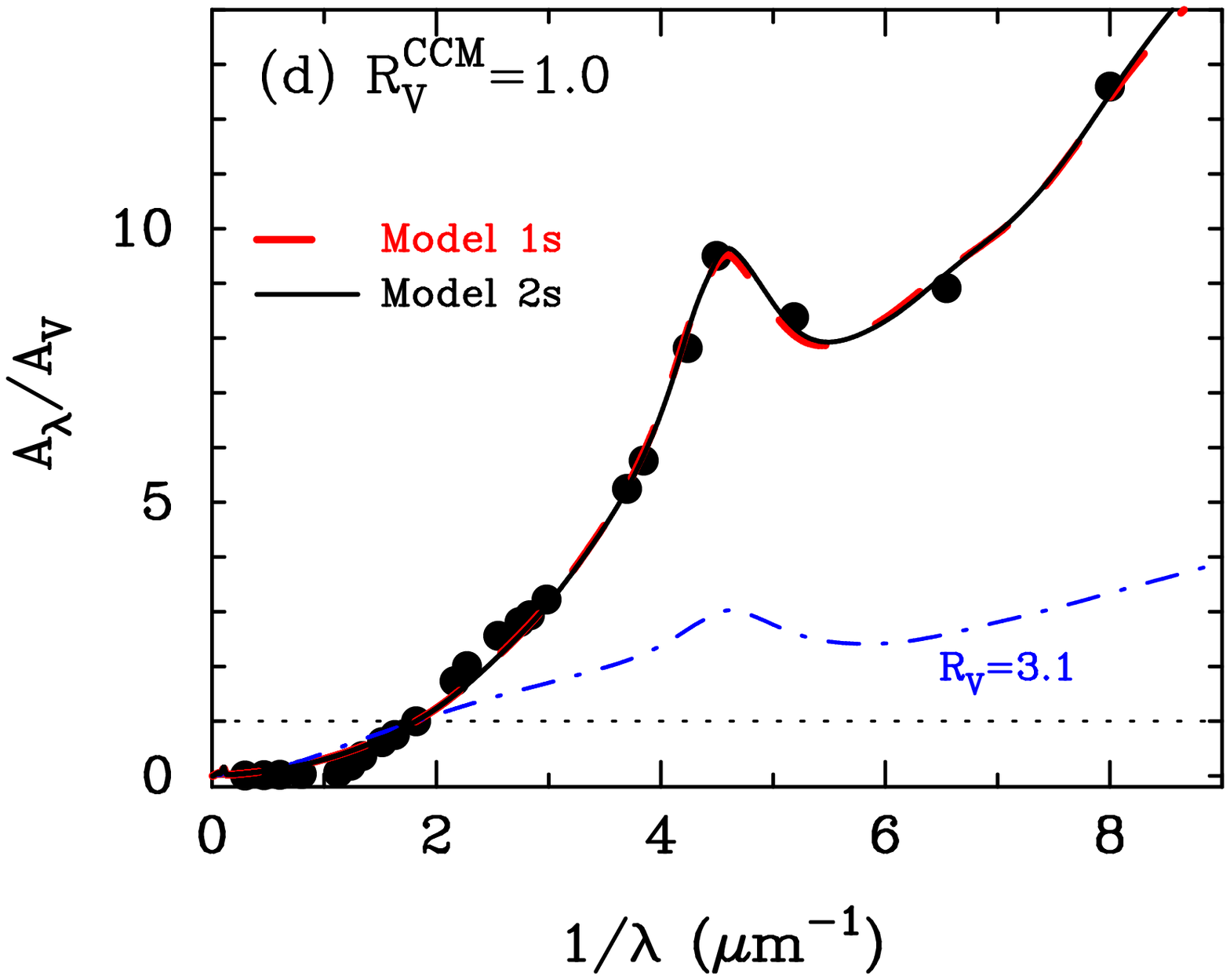}
\end{minipage}
\end{tabular}
\caption{Extinction curves calculated from the best-fit parameters 
in Model 1s (dashed lines) and Model 5s (solid lines) with lognormal 
size distributions for (a) $R_V^{\rm CCM} = 3.1$ (b) $2.0$, (c) $1.5$, 
and (d) $1.0$.
The filled circles are the extinction data at the reference wavelengths,
derived from the CCM formula for each $R_V^{\rm CCM}$ value.
For the panels (b)--(d), the extinction curve derived from Model 1 with
the power-law size distribution for $R_V^{\rm CCM} = 3.1$ is shown by 
the dot-dashed line for reference.
\label{fig:lognormal}}
\end{figure*}

\subsection{Extinction curves from dust models with lognormal size distributions} 
\label{sec:lognormal}

In this section, we examine whether the dust models with lognormal 
size distributions can lead to the good fit to the CCM curves with
$R_V^{\rm CCM} =$ 1.0--3.1.
As mentioned in Section \ref{subsec:dustmodel}, the lognormal-like 
size distributions have been suggested as those of dust supplied 
from CCSNe and AGB stars.
\citet{hir15} adopted the standard deviation of $\gamma_j = 0.49$ 
for the lognormal size distribution of dust from these stellar sources.
Following this study, we first consider the case with 
$\gamma_j = 0.5$, treating $a_{\rm 0, gra}$, $a_{\rm 0, sil}$,
and $f_{\rm gs}$ as parameters (referred to as Model 1s).
We note that $\gamma_j = 0.5$ presents a well peaked distribution, 
so such lognormal distributions would be useful to seek the 
characteristic grain radius responsible for the measured extinction 
curves.

The best-fit parameters from the fitting calculations for lognormal grain 
size distributions are given in Table \ref{tab:fitres1}.
For a fixed value of $\gamma_j = 0.5$, the best-fits are obtained with 
$a_{0, j} \simeq 0.03$ $\mu$m (for $R_V^{\rm CCM} =$ 3.1) down 
to $a_{0, j} \simeq 0.01$ $\mu$m (for $R_V^{\rm CCM} =$ 1.0), 
indicating that the characteristic grain radius decreases as the 
extinction curve becomes steeper.
However, as shown by dashed lines in Figure \ref{fig:lognormal}, the 
matches are poor, especially for $R_V^{\rm CCM} =$ 3.1, 2.0, and 1.5.
Hence,the lognormal distribution with the 
standard deviation as low as $\gamma_j = 0.5$ is not likely to be 
suitable for managing these steep CCM curves.

\begin{figure*}
\begin{tabular}{cc}
\begin{minipage}[t]{0.45\hsize}
\includegraphics[width=1\textwidth]{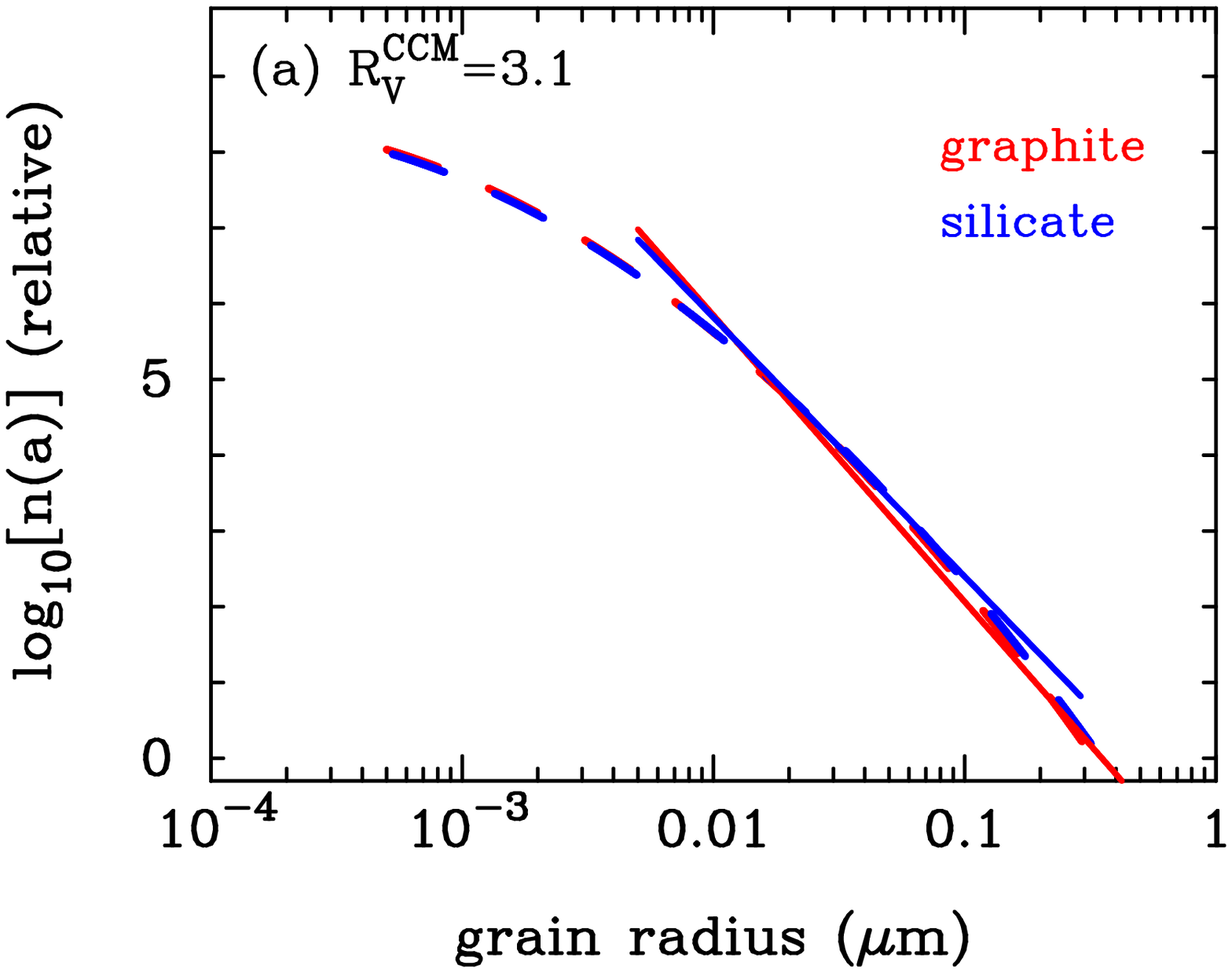}
\end{minipage} &
\begin{minipage}[t]{0.45\hsize}
\includegraphics[width=1\textwidth]{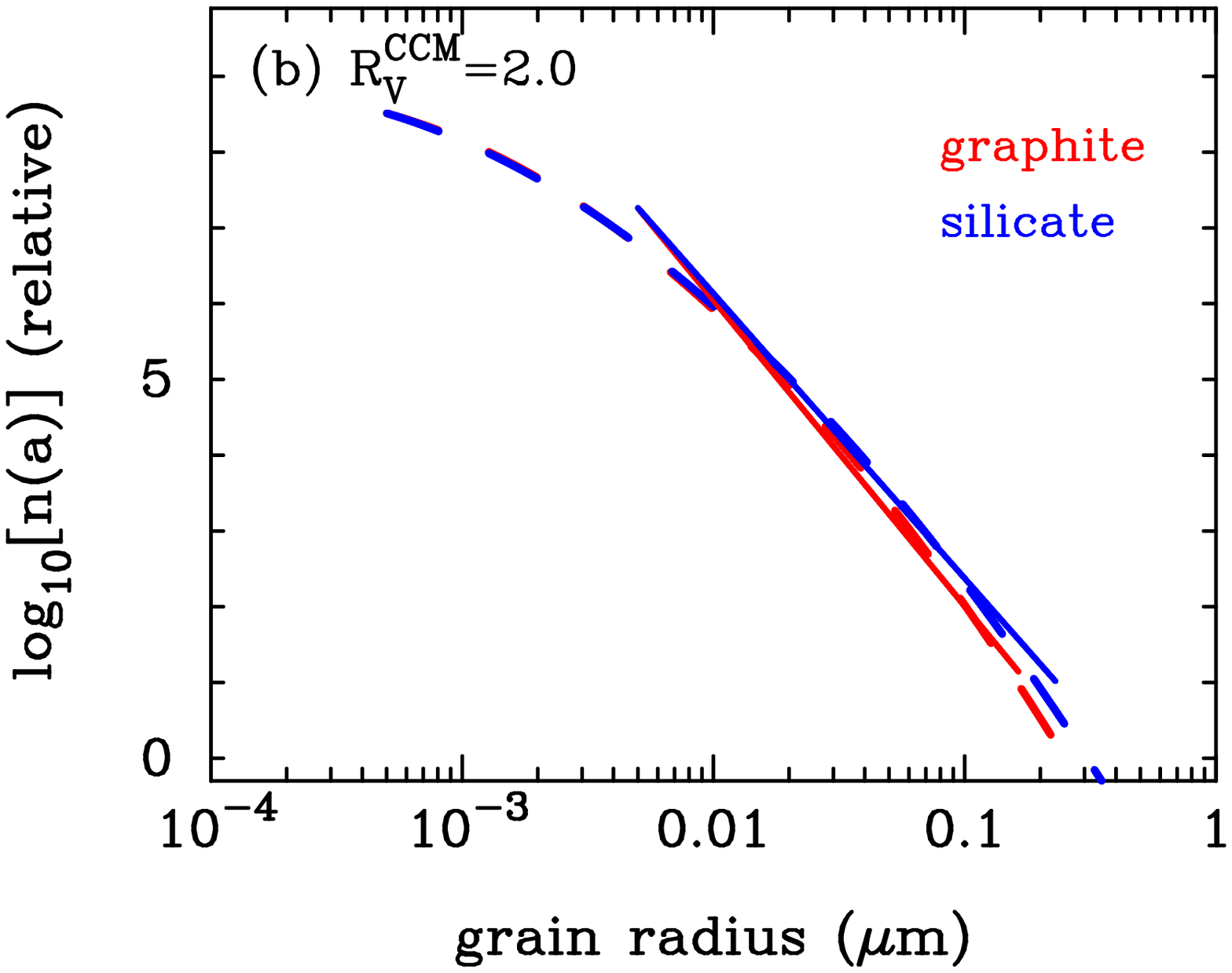}
\end{minipage}
\end{tabular}
\begin{tabular}{cc}
\begin{minipage}[t]{0.45\hsize}
\includegraphics[width=1\textwidth]{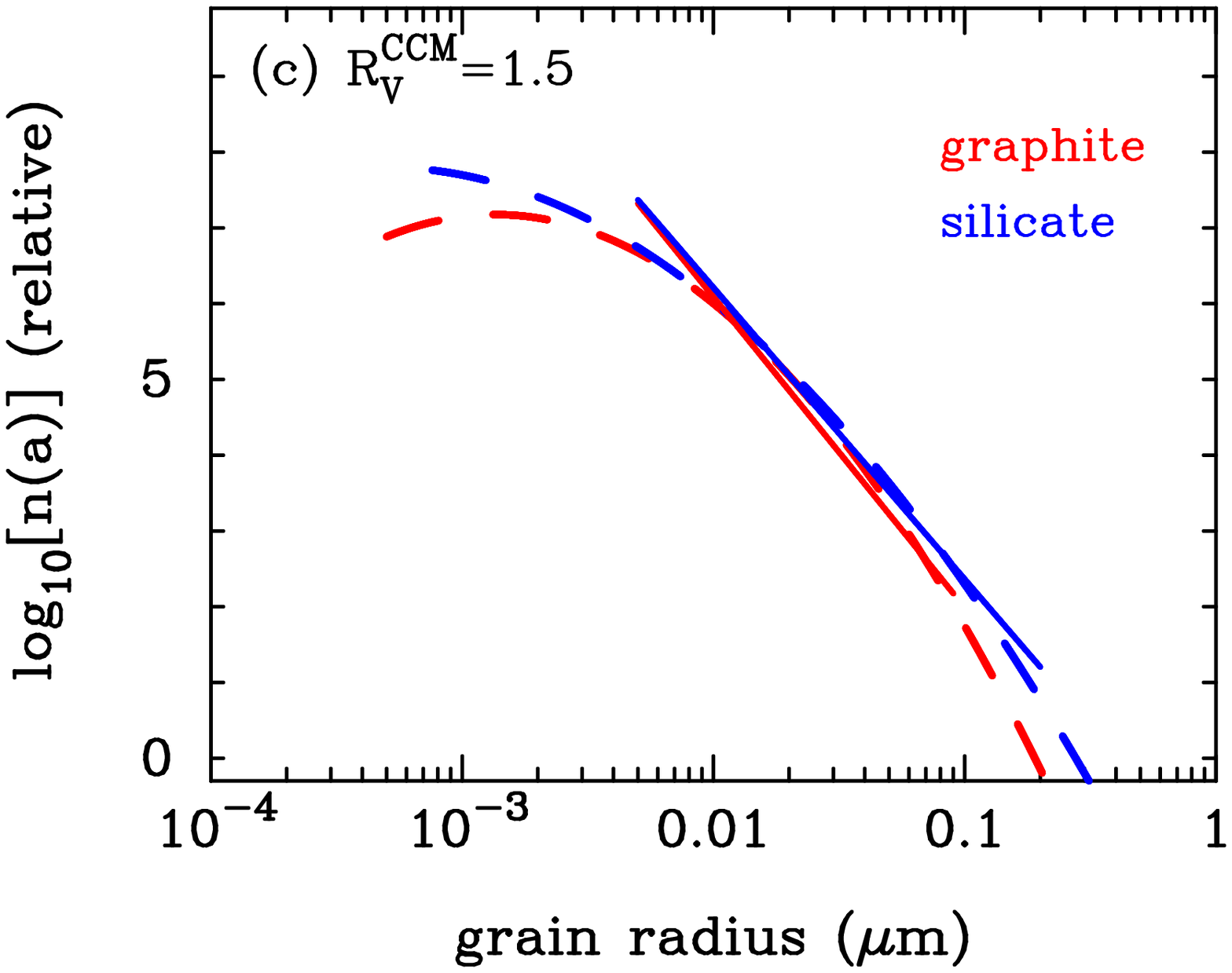}
\end{minipage} &
\begin{minipage}[t]{0.45\hsize}
\includegraphics[width=1\textwidth]{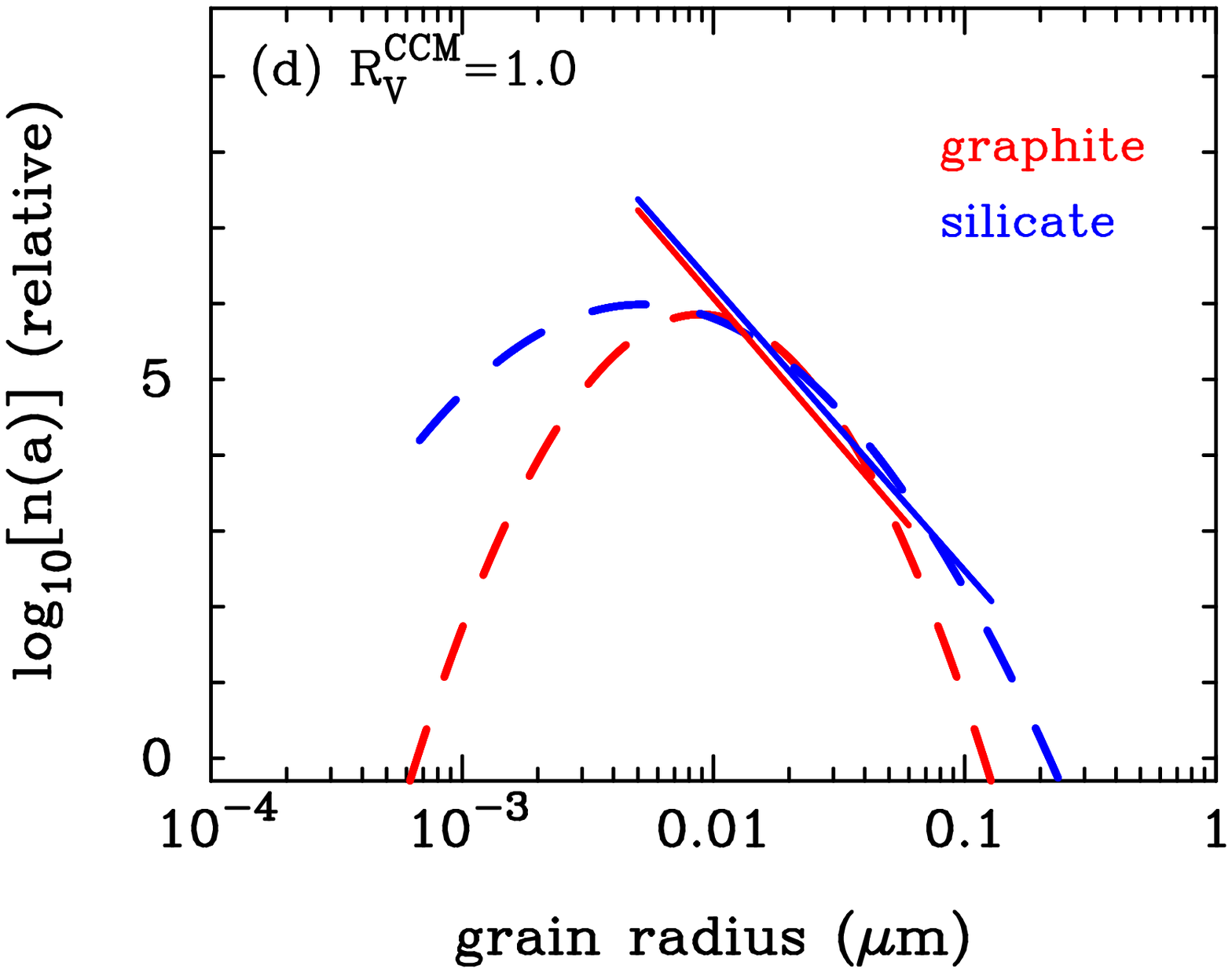}
\end{minipage}
\end{tabular}
\caption{Grain size distributions that lead to the best-fit to the 
extinction curves with (a) $R_V^{\rm CCM} = 3.1$ (b) $2.0$, 
(c) $1.5$, and (d) $1.0$.
In each panel, the solid lines show the power-law size distributions 
obtained from Model 5, while the dashed lines indicate the results from 
lognormal size distributions from Model 5s.
Graphite and silicate are drawn in red and blue, respectively.
\label{fig:sizedistribution}}
\end{figure*}

On the other hand, if we treat all of the five quantities 
($f_{\rm gs}$, $a_{\rm 0, gra}$, $ a_{\rm 0, sil}$, $\gamma_{\rm gra}$, 
and $\gamma_{\rm sil}$) involved in the lognormal distribution as free 
parameters  (referred to as Model 5s), the nice fits are obtained 
(solid lines in Figure \ref{fig:lognormal});
in the case of Model 5s, $a_{0, j} \le \sim 3 \times 10^{-3}$ $\mu$m 
and $\gamma_j \simeq 1.0$ are necessary for producing the good 
fits to the extinction curves with $R_V^{\rm CCM} =$ 3.1, 2.0, 
and 1.5.
Note that the size distribution obtained from such a small 
$a_{0, j}$ and a large $\gamma_j$ can be viewed as an 
exponential-like distribution rather than a lognormal distribution 
with a sharp peak (see Figure \ref{fig:sizedistribution}).
For $R_V^{\rm CCM} =$ 1.0, a peaked lognormal distribution 
seems to yield the entire agreement with the extinction data, 
although the match is not necessarily sufficient in optical and 
near-infrared regions.

The inspection of Figure \ref{fig:sizedistribution} allows us to 
point out that the size distributions offering the best-fits for 
Model 5s have well similar slopes to the corresponding power-law 
distributions at the radii between $\simeq$0.01 $\mu$m and 
$\simeq$0.2 $\mu$m.
Then, we see if this similarity also holds with regard to the 
average radii given as
\begin{eqnarray}
\langle a_{{\rm ave}, j}^m \rangle = \left( 
\frac{\int_{a_{{\rm min}, j}}^{a_{{\rm max}, j}} a^m n(a) da}
{\int_{a_{{\rm min}, j}}^{a_{{\rm max}, j}} n(a) da} \right)^{\frac{1}{m}} 
~~~~ (m=1, 2, 3).
\label{eq:averadi}
\end{eqnarray}
Table \ref{tab:averadi} presents the average radii from the 
best-fit parameters for some of the dust models in this study.
We can see that $\langle a_{{\rm ave}, j}^m \rangle$ from Models 
5 and 5s do not coincide at all, despite the fact that their size
distributions highly resemble at $a \simeq$ 0.01--0.2 $\mu$m.
For $R_V^{\rm CCM} =$ 3.1, 2.0, and 1.5, 
$\langle a_{{\rm ave}, j}^m \rangle$ from Model 5s are always smaller
than those from Model 5.
This is because $a_{{\rm min}, j}$ (and $a_{0, j}$) in Model 5s is 
much smaller than $a_{{\rm min}, j} = 0.005$ $\mu$m in Model 5.
Meanwhile, $\langle a_{{\rm ave}, j}^m \rangle$ is higher in Model 
5s for $R_V^{\rm CCM} =$ 1.0 due to the lack of grains smaller than
$\simeq$0.01 $\mu$m.
Hence, it would not be proper to invoke the average radius when 
the properties of dust are discussed in the context of extinction 
curves.
Indeed, the radii of grains which mainly contribute to extinction 
are different at different wavelengths, so the wavelength-dependence 
of extinction could not be described only with a single characteristic 
grain radius.

We have revealed that the lognormal size distribution can lead to 
good fits to the CCM curves with $R_V^{\rm CCM} =$ 1.0--3.1 
by adopting the appropriate $a_{0, j}$ and $\gamma_j$.
However, the required values of $a_{0, j}$ is below 0.01 $\mu$m, 
which is much smaller than the typical radii (0.1--1.0 $\mu$m) of 
dust expected from CCSNe and AGB stars.
In fact, dust grains from CCSNe could cause the flat extinction 
curves because of their relatively large radii \citep{hir08}. 
We have also noticed that the average radius could not be a 
good quantity to characterize the size distribution of dust that 
reproduces the extinction curves.

\begin{table*}
\caption{Average radii $\langle a_{{\rm ave}, j}^m \rangle$ of 
graphite and silicate grains in units of $\mu$m for Models 1 and 5 in 
Table \ref{tab:dustmodelpl} and for Model 5s in Table 
\ref{tab:dustmodelln}.} 
\label{tab:averadi}
\begin{center}
\begin{tabular}{@{}lcccccc} \hline
Dust model & & graphite &  &  & silicate & \\
 & $m=1$ & $m=2$ & $m=3$ & $m=1$ & $m=2$ & $m=3$ 
\\ \hline \hline
\multicolumn{7}{c}{$R_V^{\rm CCM} = 3.1$}  \\ \hline
Model 1 & 0.0083 & 0.0103 & 0.0155 & 0.0083 & 0.0103 & 0.0155 \\
Model 2 & 0.0082 & 0.0102 & 0.0152 & 0.0082 & 0.0102 & 0.0152 \\
Model 4 & 0.0082 & 0.0100 & 0.0150 & 0.0082 & 0.0100 & 0.0150 \\ 
Model 5 & 0.0078 & 0.0093 & 0.0139 & 0.0085 & 0.0109 & 0.0171 \\ 
Model 5s & 0.0022 & 0.0041 & 0.0087 & 0.0022 & 0.0043 & 0.0095 \\ 
\hline \hline
\multicolumn{7}{c}{$R_V^{\rm CCM} = 2.0$}  \\ \hline
Model 1 & 0.0083 & 0.0100 & 0.0138 & 0.0083 & 0.0100 & 0.0138 \\
Model 2 & 0.0074 & 0.0085 & 0.0111 & 0.0074 & 0.0085 & 0.0111 \\
Model 4 & 0.0077 & 0.0090 & 0.0119 & 0.0077 & 0.0090 & 0.0119 \\
Model 5 & 0.0074 & 0.0084 & 0.0110 & 0.0078 & 0.0093 & 0.0129 \\ 
Model 5s & 0.0019 & 0.0031 & 0.0056 & 0.0020 & 0.0034 & 0.0066 \\ 
\hline \hline
\multicolumn{7}{c}{$R_V^{\rm CCM} = 1.5$}  \\ \hline
Model 1 & 0.0082 & 0.0098 & 0.0128 & 0.0082 & 0.0098 & 0.0128 \\
Model 2 & 0.0071 & 0.0078 & 0.0094 & 0.0071 & 0.0078 & 0.0094 \\
Model 4 & 0.0079 & 0.0093 & 0.0120 & 0.0079 & 0.0093 & 0.0120 \\
Model 5 & 0.0074 & 0.0082 & 0.0099 & 0.0077 & 0.0090 & 0.0121 \\ 
Model 5s & 0.0042 & 0.0060 & 0.0087 & 0.0030 & 0.0049 & 0.0084 \\ 
\hline \hline
\multicolumn{7}{c}{$R_V^{\rm CCM} = 1.0$}  \\ \hline
Model 1 & 0.0081 & 0.0094 & 0.0114 & 0.0081 & 0.0094 & 0.114 \\
Model 2 & 0.0066 & 0.0070 & 0.0077 & 0.0066 & 0.0070 & 0.0077 \\
Model 4 & 0.0085 & 0.0100 & 0.0122 & 0.0085 & 0.0100 & 0.0122 \\
Model 5 & 0.0076 & 0.0086 & 0.0102 & 0.0078 & 0.0091 & 0.0119 \\ 
Model 5s & 0.0129 & 0.0146 & 0.0166 & 0.0111 & 0.0143 & 0.0184 \\ 
\hline
\end{tabular}
\end{center}
\end{table*}

\subsection{Allowed ranges of $a_{\rm max}$, $q$ and $f_{\rm gs}$ for 
power-law size distribution} 
\label{sec:allowedrange}

In Section \ref{sec:powerlaw}, we have shown that the CCM extinction 
curves with $R_V^{\rm CCM} =$ 1.0--3.1 can be reasonably fitted by 
the simplest power-law dust model with $q = 3.5$ through taking an 
appropriate set of $a_{\rm max}$ and $f_{\rm gs}$.
We found that, for Model 1,  the optimum maximum cut-off radius 
$a_{\rm max}$ decreases from $a_{\rm max} = 0.24$ $\mu$m for 
$R_V = 3.1$ down to $a_{\rm max} = 0.057$ $\mu$m for 
$R_V = 1.0$, with a range of $f_{\rm gs} =$ 0.45--0.6.
However, the best-fit value is not a unique solution;
there should be other combinations of $a_{\rm max}$ and $f_{\rm gs}$ 
that still yield reasonable fits to the extinction data.
Therefore, given that there are some uncertainties on the data of 
extinction curves, it should be inspected what extent of the 
change in $a_{\rm max}$ and $f_{\rm gs}$ is allowable.

In order to quantify the allowed ranges of $a_{\rm max}$ and 
$f_{\rm gs}$, we introduce the average extinction uncertainty 
$\tilde{\sigma}$, defined as 
$\tilde{\sigma} = \sum \sigma_i / N_{\rm data}$, where $\sigma_i$ 
are the uncertainties of extinction data at the reference wavelengths
$\lambda_i$.
Then, if the dispersion $\chi_1$ calculated for a given combination 
of $a_{\rm max}$ and $f_{\rm gs}$ is smaller than $\tilde{\sigma}$, 
we consider it as reproducing the extinction curves within the 
1$\sigma$ errors.
Using the values of $\sigma_i$ in Table \ref{tab:refwl} leads to
the average uncertainty of $\tilde{\sigma} = 0.115$.
It should be kept in mind that some of $\sigma_i$ in Table 
\ref{tab:refwl} are inferred from those at the closest reference 
wavelengths, so the absolute value of $\tilde{\sigma}$ is quite arbitrary.
Nevertheless, the introduction of such a criterion for the dispersion 
would give a meaningiful indication for the allowed range of 
$a_{\rm max}$ and $f_{\rm gs}$.

\begin{figure}
\includegraphics[width=0.45\textwidth]{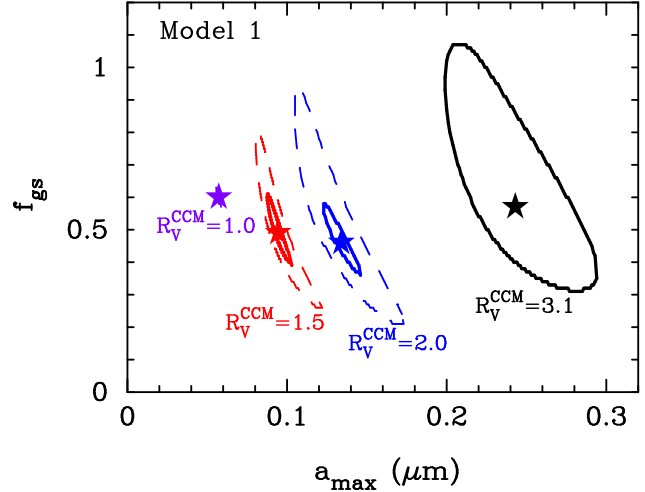}
\caption{Contour plots showing the allowed regions of $a_{\rm max}$ 
and $f_{\rm gs}$ with which Model 1 can yield fits to the CCM extinction 
curves within the 1$\sigma$ errors (solid lines) for 
$R_V^{\rm CCM} = $ 1.5, 2.0, and 3.1.
For $R_V^{\rm CCM} =$ 1.0, 1.5, and 2.0, the contours within 
the 2$\sigma$ errors are also drawn by the dashed lines.
The best-fit combination of $a_{\rm max}$ and $f_{\rm gs}$ is marked by 
the filled star for each $R_V^{\rm CCM}$.
Note that the size of 2$\sigma$ contour for $R_V^{\rm CCM} = $ 1.0 
is comparable with that of the star symbol. 
\label{fig:contour1}}
\end{figure}

Figure \ref{fig:contour1} shows the contours within which the 
combinations of $a_{\rm max}$ and $f_{\rm gs}$ satisfy the 1$\sigma$ 
condition $\chi_1 \le \tilde{\sigma}$ for $R_V^{\rm CCM} =$ 
3.1, 2.0, and 1.5.
For $R_V^{\rm CCM} = 1.0$, there is no combination of $a_{\rm max}$ 
and $f_{\rm gs}$ within the 1$\sigma$ error.
Thus, we plot the 2$\sigma$ range (that is, combinations of 
$a_{\rm max}$ and $f_{\rm gs}$ with $\chi_1 \le 2 \tilde{\sigma}$) 
for $R_V^{\rm CCM} = 1.0$ (we also show the 2$\sigma$ contours for 
$R_V^{\rm CCM} =$ 2.0 and 1.5).
We observe that, for each $R_V^{\rm CCM}$, a higher $f_{\rm gs}$ 
is needed for a lower $a_{\rm max}$ to produce fits within the 
1$\sigma$ errors for $R_V^{\rm CCM} = $1.5--3.1.
For $R_V^{\rm CCM} = 3.1$, the 1$\sigma$ ranges of 
$a_{\rm max}$ and $f_{\rm gs}$ are 0.199 $\mu$m $\le a_{\rm max} \le$ 
0.294 $\mu$m and 0.31 $\le f_{\rm gs} \le$ 1.07,
whereas the allowed ranges of $a_{\rm max}$ lie just around its optimum 
values for $R_V^{\rm CCM} \le 2.0$, keeping the trend that
$a_{\rm max}$ decreases with decreasing $R_V^{\rm CCM}$.
On the other hand, $f_{\rm gs}$ is relatively uniform with the values of 
$\simeq$0.4--0.6, regardless of $R_V^{\rm CCM}$.
This means that the fraction of graphite mass of the total dust mass 
is in a narrow range of $f_{\rm gs} / (1 + f_{\rm gs}) \simeq$ 0.3--0.4.

\begin{figure}
\includegraphics[width=0.45\textwidth]{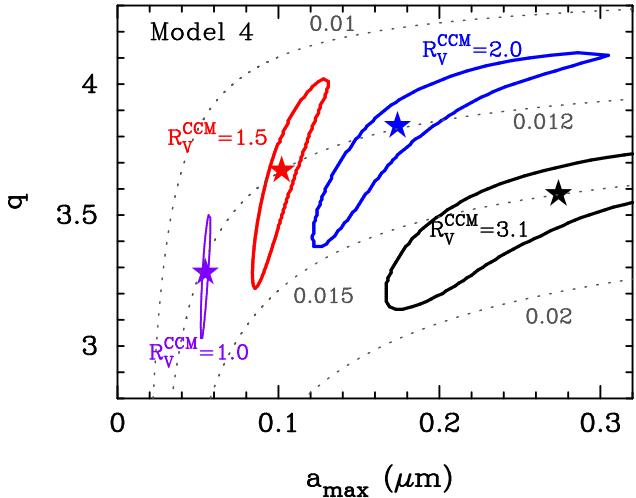}
\caption{Contour plots showing the allowed regions of $a_{\rm max}$ 
and $q$ with which Model 4 can yield fits to the CCM extinction 
curves within the 1$\sigma$ errors for $R_V^{\rm CCM} = $ 1.5, 
2.0, and 3.1 (thick solid lines).
For $R_V^{\rm CCM} =$ 1.0, the contour within the 2$\sigma$ errors 
is drawn by thin solid line.
The best-fit combination of $a_{\rm max}$ and $q$ is marked by 
the filled star for each $R_V^{\rm CCM}$.
The dotted curves indicate the trajectories of specific average radii
$\langle a_{{\rm ave}j}^3 \rangle$ calculated from 
Equation (\ref{eq:averadi}) adopting $m = 3$:
$\langle a_{{\rm ave}j}^3 \rangle = $ 0.01, 0.012, 0.015, and 0.02 
$\mu$m from top to bottom.
\label{fig:contour2}}
\end{figure}

Next we consider the variation of the power-law index $q$.
As demonstrated in Section \ref{sec:powerlaw}, Model 2, in 
which $a_{\rm max}$ is fixed as 0.25 $\mu$m, cannot reproduce 
the CCM curves with $R_V^{\rm CCM} =$ 1.5, and 1.0; 
in these cases, the dispersion $\chi_1$ is larger than 
$2 \tilde{\sigma} = 0.23$ even for their best-fit combinations 
of $q$ and $f_{\rm gs}$ (see Table \ref{tab:fitres}).
This implies that the change in $a_{\rm max}$ must be essential
for the reproduction of the steep extinction curves with 
$R_V^{\rm CCM} \le 1.5$.
Figure \ref{fig:contour2} presents the $1 \sigma$ 
($2 \sigma$) ranges of combination of $a_{\rm max}$ and $q$ 
for $R_V^{\rm CCM} =$ 3.1, 2.0, and 1.5 ($R_V^{\rm CCM} =$ 1.0)
under the assumption that graphite and silicate have the same 
size distribution
(i.e., $a_{\rm max} = a_{\rm max, gra} = a_{\rm max, sil}$ and 
$q = q_{\rm gra} = q_{\rm sil}$ which corresponds to Model 4).
The contours show that, for a given value of $R_V^{\rm CCM}$, 
a higher $q$ is needed for a higher 
$a_{\rm max}$.\footnote{\citet{noz13} showed that, to reproduce
the average MW extinction curve, the ranges of 
$q = 3.5 \pm 0.2$ and $a_{\rm max} = 0.24 \pm 0.05$ $\mu$m 
are demanded, which are narrower than that given in 
Figure \ref{fig:contour2}.
They took into account the constraints of elemental abundances,
and adopted the definition different from that in this study for 
the $1 \sigma$ error of the extinction curve.}
However, the power index does not increase with going from 
$R_V^{\rm CCM} =2.0$ down to 1.0;
$q$ is confined within the range of $q=$ 3--4 for any value of 
$R_V^{\rm CCM}$ considered in this work, while there is a 
clear tendency that $a_{\rm max}$ decreases with decreasing 
$R_V^{\rm CCM}$.
This indicates that the reduction in $a_{\rm max}$ is preferable to 
the enhancement in $q$ for describing the highly steep 
extinction curves.
We also note that, for the cases of $R_V^{\rm CCM} =$ 2.0, 1.5, 
and 1.0, the best-fit combinations of $a_{\rm max}$ and $q$ in 
Model 4 lead to the similar average radii of 
$\langle a_{{\rm ave}, j}^3 \rangle \simeq 0.012$ $\mu$m.
This is an additional indication that the average radius is not
a good measure of size distributions that feature the extinction 
curves.

In summary, the variation of the extinction curves is mostly 
described by the change in $a_{\rm max}$ in the context of the 
power-law size distribution.
For $R_V^{\rm CCM} =$ 1.0--3.1, the range of $q =$ 3--4 
reasonably fits to the CCM curves.
The mass ratio of graphite to silicate is constrained to be 
$0.4 \le f_{\rm gs} \le 0.6$, which interestingly well agrees with 
the range $0.3 \le f_{\rm gs} \le 0.7$ estimated from the analysis 
of the average extinction curve and abundance constraints in the 
MW \citep{noz13}.
This suggests that the composition of dust toward lines of sight 
with $R_V^{\rm CCM} \le 2.0$ would not be largely 
different from that in the MW.

\section{Discussions} 
\label{sec:discussion}

\subsection{Dependence of $R_V$ on $a_{\rm max}$} 
\label{sec:rvamax}

We have demonstrated that the two-component model of graphite and
silicate with power-law size distributions is a competent dust model that
can explain the systematic behaviors of extinction curves for a wide 
variety of $R_V^{\rm CCM}$.
However, the $R_V^{\rm mod}$ calculated from our dust models does
not match the $R_V^{\rm CCM}$ that is referred to in deriving 
the data of the extinction curve;
for all the power-law dust models considered in this study, 
$R_V^{\rm mod}$ is higher than $R_V^{\rm CCM}$ as seen from 
Table \ref{tab:fitres}.
This indicates that our dust model can account for the overall shape 
of extinction curves but cannot accurately reproduce the extinction 
ratio between specific wavelengths.

\begin{figure}
\includegraphics[width=0.45\textwidth]{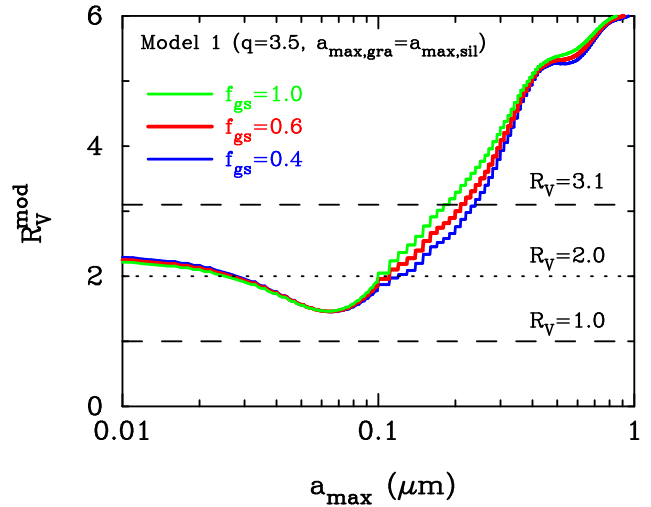}
\caption{Dependence of $R_V^{\rm mod}$ on $a_{\rm max}$ calculated 
from Model 1 with $q = 3.5$ and 
$a_{\rm max} = a_{\rm max, gra} =a_{\rm max, sil}$
for $f_{\rm gs} =$ 0.4, 0.6, and 1.0.
The minimum cut-off radii are fixed to $a_{\rm min} =$ 0.005 $\mu$m.
The two dashed horizontal lines represent $R_V = $ 
3.1 and 1.0, and the dotted line exhibits $R_V = $ 2.0.
\label{fig:rvamax}}
\end{figure}

Figure \ref{fig:rvamax} depicts the $R_V^{\rm mod}$ calculated from 
Model 1 as a function of $a_{\rm max}$ for $f_{\rm gs} =$ 0.4, 0.6, 
and 1.0.
We can see that the dependence of $R_V^{\rm mod}$ on $f_{\rm gs}$
is weak, although there appears slight difference at $a_{\rm max} = $
0.1--0.4 $\mu$m. 
More importantly, as $a_{\rm max}$ increases, $R_V^{\rm mod}$ 
decreases at $a_{\rm max} \le 0.06$ $\mu$m and increases at 
$a_{\rm max} \ge 0.07$ $\mu$m, having the minimum value of 
$R_V^{\rm mod} \simeq 1.46$ at $a_{\rm max} \simeq 0.065$ $\mu$m
for any value of $f_{\rm gs}$.
We have confirmed that this dependence of $R_V^{\rm mod}$ on
$a_{\rm max}$ is not sensititve to the minimum cut-off radius 
$a_{\rm min}$ as long as $a_{\rm min} \le 0.01$ $\mu$m.
Thus, if the measured values of $R_V < 1.4$ are real, they could
not be interpreted by the simple dust model considered here.
The unfeasibility of such low $R_V$, as well as the descrepancy 
between $R_V^{\rm mod}$ and $R_V^{\rm CCM}$, implies that more 
sophisticated dust models which adopt more fine-tuned size distribution 
function and involve additional components other than graphite and 
silicate are necessary for consistently explaining the shape of extinction 
curve and the value of $R_V$.

We also note that the measured value of $R_V$ is unlikely to be 
a good probe for constraining the properties of interstellar dust.
For example, as seen in Figure \ref{fig:rvamax}, there are two
possible values of $a_{\rm max}$ ($a_{\rm max} \simeq$ 0.026 $\mu$m 
and 0.11 $\mu$m) that realize $R_V = 2.0$.
This suggests that the properties of dust cannot be necessarily 
determined uniquely only from $R_V$ being the ratio of extinction in 
$V$ and $B$ bands, unless $R_V$ is an extremely low or high value.
In order to extract the information on the properties of interstellar 
dust, the extinction data over a broad range of wavelengths are 
essential, which is demonstrated in more details in the next subsection.

\subsection{Necessity of UV extinction data} 
\label{sec:nonuv}

So far, we have performed the fitting to the extinction data at all 
the reference wavelengths covering UV to near-infrared.
However, the data at UV wavelengths cannot be always acquired,
and in most cases, we have to derive the extinction law by 
relying only on the data at optical to near-infrared wavelengths.
In addition, the extinction curves in external galaxies such as Large 
and Small Magellanic Clouds do not show a remarkable 2175 
\AA~bump \citep{gor03} and cannot be described as the CCM curves.
Thus, it would be interesting to see how the results of fitting can be
changed in the cases that the UV extinction data are not taken into 
account, especially in the absence of data around the 2175 \AA~bump.

\begin{table*}
\caption{A set of the best-fit parameters obtained for some dust models
that do not consider the UV extinction data.}
\label{tab:nouv}
\begin{center}
\begin{tabular}{@{}lccccccc} \hline
Dust model$\,^a$ & $q_{\rm gra}$ & $a_{\rm max, gra}$ & $q_{\rm sil}$ & 
$a_{\rm max, sil}$ & $f_{\rm gs}$ & $\chi_1$ & $R_V^{\rm cal}$ \\
 & & ($\mu$m) &  & ($\mu$m) & & & 
\\ \hline \hline
\multicolumn{8}{c}{$R_V^{\rm CCM} = 3.1$}  \\ \hline
Model 1nb & 3.50 & 0.259 & 3.50 & 0.259 & 0.50 & 0.0224 & 3.41 \\
Model 1nu & 3.50 & 0.266 & 3.50 & 0.266 & 0.41 & 0.0234 & 3.42 \\
\hline \hline
\multicolumn{8}{c}{$R_V^{\rm CCM} = 2.0$}  \\ \hline
Model 1nb & 3.50 & 0.128 & 3.50 & 0.128 & 0.50 & 0.0605 & 2.12 \\
Model 5nb & 1.50 & 0.0729 & 3.94 & 0.441 & 0.18 & 0.0286 & 2.11 \\ 
Model 1nu & 3.50 & 0.201 & 3.50 & 0.201 & 0.16 & 0.0354 & 2.30 \\
\hline \hline
\multicolumn{8}{c}{$R_V^{\rm CCM} = 1.5$}  \\ \hline
Model 1nb & 3.50 & 0.0896 & 3.50 & 0.0896 & 0.55 & 0.0553 & 1.68 \\
Model 5nb & 2.00 & 0.0741 & 2.31 & 0.0358 & 0.63 & 0.0322 & 1.53 \\ 
Model 1nu & 3.50 & 0.0830 & 3.50 & 0.0830 & 4.00 & 0.0458 & 1.65 \\
\hline \hline
\multicolumn{8}{c}{$R_V^{\rm CCM} = 1.0$}  \\ \hline
Model 1nb & 3.50 & 0.0618 & 3.50 & 0.0618 & 0.53 & 0.215 & 1.46 \\
Model 1nu & 3.50 & 0.149 & 3.50 & 0.149 & 0.00 & 0.115 & 1.18 \\
\hline
\end{tabular}
\end{center}
$^a$Model 1nb and Model 1nu are the same as Model 1 in Table 
\ref{tab:dustmodelpl} but do not take into account, respectively, 
the five data around the 2175 \AA~bump at $1/\lambda =$ 3.0--5.2 
$\mu$m$^{-1}$ and the seven UV data at $1/\lambda <$ 3.0 
$\mu$m$^{-1}$ for the fitting calculations. 
Model 5nb is the same dust model as Model 5 in Table 
\ref{tab:dustmodelpl} but does not take the data around the UV 
bump in consideration.
\end{table*}

\begin{figure*}
\begin{tabular}{cc}
\begin{minipage}[t]{0.45\hsize}
\includegraphics[width=1\textwidth]{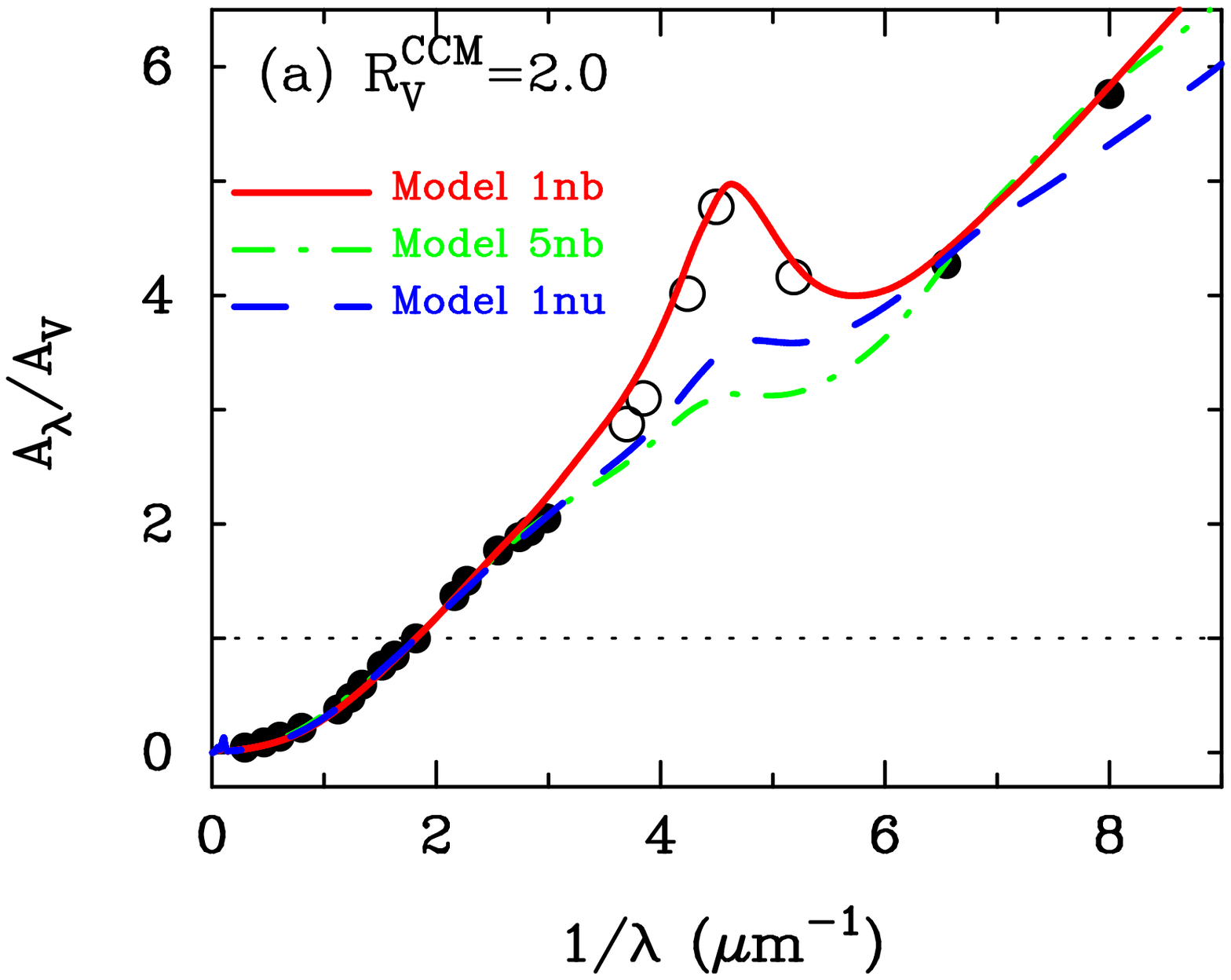}
\end{minipage} &
\begin{minipage}[t]{0.45\hsize}
\includegraphics[width=1\textwidth]{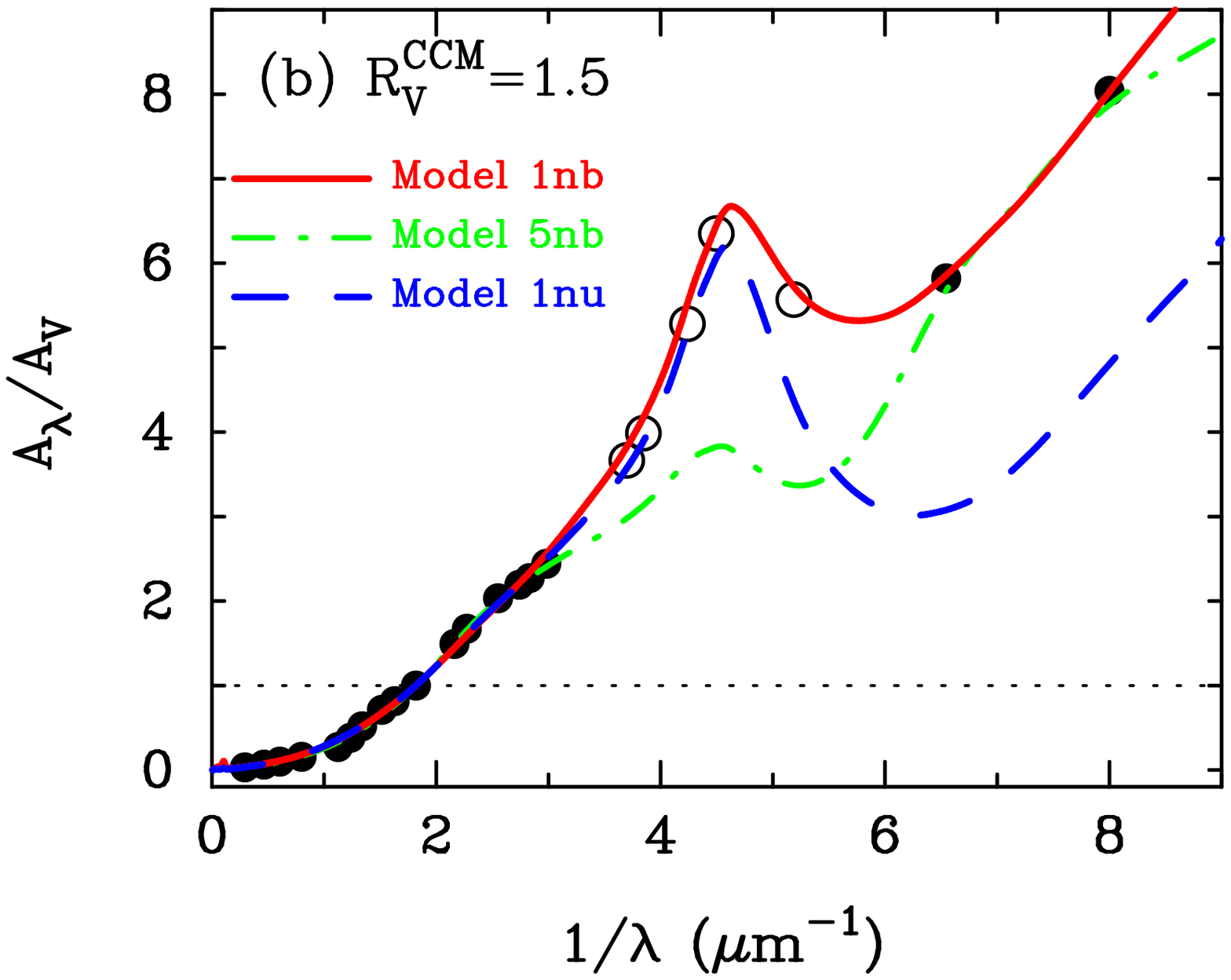}
\end{minipage}
\end{tabular}
\caption{Extinction curves for (a) $R_V^{\rm CCM} = 2.0$ and (b)
$R_V^{\rm CCM} = 1.5$, calculated from the best-fit combonations
of $a_{\rm max}$ and $f_{\rm gs}$ for Model 1nb (solid) and Model 
1nu (dashed) with $q = 3.5$ and $a_{\rm max, gra} = a_{\rm max, sil}$.
Model 1nb performs the fitting to the data of the extinction curves 
in which the five data around the UV bump at 
$1/\lambda =$ 3.0--5.2 $\mu$m$^{-1}$ are excluded, while 
Model 1nu does not include the UV data at $1/\lambda <$ 3.0 
$\mu$m$^{-1}$ for the fitting calculations.
The dot-dashed line depicts the extinction curves from Model 5nb, 
which uses the same extinction data as Model 1nb but parameterizes
all of the five quantities relevant to the power-law distribution.
The circles are the extinction data at the reference wavelengths, 
derived from the CCM formula for each $R_V^{\rm CCM}$ value,
where the open symbols signify the data disregarded for Model 1nb 
and Model 5nb. 
\label{fig:nonuv}}
\end{figure*}

From this motivation, we do the fitting calculations for the data 
sets that do not include the wavelengths ($1/\lambda =$ 
3.0--5.2 $\mu$m$^{-1}$) around the 2175 \AA~bump (but still 
include the two far-UV wavelengths at $1/\lambda =$ 6.5 and 
8.0 $\mu$m$^{-1}$).
For purposes of illustration, we here focus on the dust models 
with power-law size distributions.
As been done before, we evaluate the goodness of fitting with 
Equation (\ref{eq:dispers2}), for which $N_{\rm data} = 17$ with
the five data around the UV bump being excluded.

Table \ref{tab:nouv} presents the best-fit parameters obtained
from the fitting calculations that do not take account of the data 
around the UV bump for some dust models considered in this study.
The corresponding extinction curves are shown in 
Figure \ref{fig:nonuv}(a) and \ref{fig:nonuv}(b) for 
$R_V^{\rm CCM} = 2.0$ and 1.5, respectively.
We can see that, when we assume $q = 3.5$ and 
$a_{\rm max, gra} = a_{\rm max, sil}$ (referred to as Model 1nb), 
the best-fit combinations of $a_{\rm max}$ and $f_{\rm gs}$, 
even if the UV-bump data are not considered, are chosen
so that the resulting extinction curves have a prominent 2175 
\AA~bump like those from Model 1, which do not neglect the 
UV-bump data.
This seems that the data near 2175 \AA~may be insignificant 
in extracting the properties of dust from the measured 
extinction curves.
However, this is not true;
when we parameterize all of the five quantities ($q_{\rm gra}$, 
$a_{\rm max, gra}$, $q_{\rm sil}$, $a_{\rm max, sil}$, and $f_{\rm gs}$) 
relevant to the power-law size distribution (referred to as Model 
5nb), the best-fit models yield a much weaker 2175 \AA~bump 
for both $R_V^{\rm CCM} = 2.0$ and 1.5.
In particular, the result without the UV-bump data for 
$R_V^{\rm CCM} = 2.0$ prefers a highly silicate-dominated dust 
composition ($f_{\rm gs} = 0.18$, compared to the nominal range 
of $f_{\rm gs} =$ 0.4--0.6), indicating that the data around the
2175 \AA~bump are essential for determining the mass fraction 
of graphite.

In addition, if we do not include the data at UV wavelengths 
shorter than 0.3 $\mu$m, distinct extinction curves appear 
even in the case of $q = 3.5$ and 
$a_{\rm max, gra} = a_{\rm max, sil}$ (referred to as Model 1nu for 
which $N_{\rm data} = 15$ without the seven data in UV regions).
As seen from Figure \ref{fig:nonuv}, the extinction curves
from Model 1nu for $R_V^{\rm CCM} =$ 2.0 and 1.5 well 
reproduce the extinction data at $1/\lambda \le$ 3.0 
$\mu$m$^{-1}$.
However, at $1/\lambda \ge 3.0$ $\mu$m$^{-1}$, these extinction 
curves do not resemble to their corresponding ones from Model 1;
the UV extinction curve for $R_V^{\rm CCM} =$ 2.0 shows a less 
prominent 2175 \AA~bump because of a smaller abundance of 
graphite ($f_{\rm gs} = 0.16$), whereas the best fit for 
$R_V^{\rm CCM} =$ 1.5 requires a very high abundance of 
graphite ($f_{\rm gs} \ge 4.0$), resulting in an extremely 
conspicuous UV bump.

These analyses clearly demonstrate that, despite using the same 
extinction data at optical to near-infrared wavelengths, the 
different results come out depending on whether the UV extinction 
data are taken into account or not.
Therefore, we conclude that the extinction data at UV wavelengths 
are crucial for reliablely assessing the composition and size distribution 
of interstellar dust from the measured extinction curves.

\section{Conclusion}\label{sec:conclusion}

We have investigated the properties of interstellar dust responsible
for the peculiar extinction laws with $R_V =$ 1.0--2.0 measured 
toward Type Ia supernovae (SNe Ia).
We perform the fitting calculations to the measured extinction 
curves, for which we adopt the extinction values at the reference 
wavelengths derived with the empirical one-parameter formula 
by \citet{car89}.
As a dust model for calculating the extinction curves, we consider
a two-component model composed of graphite and silicate with 
power-law and lognormal size distributions.

We first confirm that our dust model can reproduce the extinction 
curves with $R_V^{\rm CCM} = 3.1$ that is taken to be a typical 
value in the Milky Way (MW).
Then, we find that even the simplest dust model with the power 
index of $q = 3.5$ can account for the entire shapes of steep 
extinction curves described by $R_V^{\rm CCM} = 2.0$, $1.5$, and 
$1.0$ with appropriate combinations of the maximum cut-off radius 
$a_{\rm max}$ and mass ratio of graphite to silicate $f_{\rm gs}$.
In particular, $a_{\rm max}$ is found to be an important quantity to 
describe the variety of extinction curves, and decreases from 
$\simeq$0.24 $\mu$m for $R_V^{\rm CCM} = 3.1$ down to 
$\simeq$0.06 $\mu$m for $R_V^{\rm CCM} = 1.0$.
On the other hand, $f_{\rm gs}$ takes a relatively narrow range of
$\simeq$0.4--0.6, indicating the mass ratio of graphite to silicate is 
not changed dramatically for different $R_V^{\rm CCM}$.

We have demonstrated that the lognormal grain size distribution
can also work well in reproducing the CCM curves with 
$R_V^{\rm CCM} =$ 1.0--3.1 by taking the very small 
characteristic radius $a_{0, j}$ and relatively large standard 
deviation $\gamma_j$ of the distribution.
Furthermore, from the comparison of the average grain radii 
between the best-fit power-law and lognormal size distributions, 
we suggest that the average radius is not a proper quantity as
representing the properties of dust that explains the extinction 
curves.
Finally, we point out that the extinction data at a limited range of 
wavelengths, such as a single value of $R_V$, do not allow us 
to uniquely determine the properties of dust, and that the 
extinction data over a wide range of wavelengths including the UV 
data are essential for constraining the composition and size 
distribution of interstellar dust.

Our main conclusion is that, in order to explain the steep extinction 
curves suggested for SNe Ia, the size distribution of interstellar dust 
in their host galaxies is biased to small sizes, compared to that in 
the MW.
Since SNe Ia are known to happen in any types of galaxies, this 
implies that large grains ($a \ge$ $\sim$ 0.1 $\mu$m) are lacking 
in galaxies other than the MW.
Why are the sizes of interstellar dust so small in external galaxies? 
This may presuppose that the evolution history of the MW is
somehow different from the other galaxies so that the interstellar 
dust in the MW has the exceptional properties.
There is also a possibility that the measured extinction curves do 
not reflect pure extinction but are distorted by other effects such 
as contamination of scattered lights at shorter wavelengths.
\citet{nag16} showed that small silicate grains and/or polycyclic 
aromatic hydrocarbons (PAHs) are needed for producing a low 
$R_V$ through multiple scattering.
Hence, this possibility may require that the host galaxies of SNe 
Ia contain abundant small silicate grains and PAHs.
Another possibility is that something is amiss in the process of 
extracting the extinction curves from the observations of SNe Ia.
Given that SNe Ia are ideal objects to measure the extinction 
curves in external galaxies, these subjects should be addressed 
from both observational and theoretical points of view.

\section*{Acknowledgment}

We thank the anonymous referees for their careful reading and
useful comments, which highly improved the paper.
I also thank Takashi Kozasa for kindly providing me with the 
computational resources to do the fitting calculations in this study.
This work was achieved using the grant of Research Assembly 
supported by the Research Coordination Committee, National 
Astronomical Observatory of Japan (NAOJ).
This work is supported in part by the JSPS Grant-in-Aid for
Scientific Research (26400223).




\end{document}